\documentclass[usenatbib]{mn2e}
\usepackage[letterpaper,margin=0.8in]{geometry}
\usepackage{natbibmnfix,fixltx2e}
\usepackage{astrojournals}
\usepackage{mathtools}
\usepackage{amsmath}
\usepackage{txfonts}
\usepackage{graphicx}
\usepackage{microtype}
\usepackage{xspace}
\usepackage{xcolor}
\usepackage[utf8]{inputenc}
\usepackage{threeparttable}
\usepackage[T1]{fontenc}
\usepackage{aecompl}

\def\simgt{\hbox{\,\rlap{\raise 0.425ex\hbox{$>$}}\lower 0.65ex\hbox{$\sim$}\,}}
\def\simlt{\hbox{\,\rlap{\raise 0.425ex\hbox{$<$}}\lower 0.65ex\hbox{$\sim$}\,}}
\def\grale{\textsc{Grale}}
\def\lenstool{\textsc{Lenstool}}

\renewcommand{\vec}[1]{\boldsymbol{#1}} % This should replace the arrows in vector notation with a bold symbol
\date{}

\pagerange{\pageref{firstpage}--\pageref{lastpage}}
\pubyear{2016}

\begin{document}
\label{firstpage}

\title[Lens Models Under the Microscope]
{Lens Models Under the Microscope: Comparison of Hubble Frontier Field Cluster Magnification Maps}

\author[J. Priewe et al.]{Jett Priewe$^{1}$, Liliya L.R. Williams$^{1}$, Jori Liesenborgs$^{2}$, Dan Coe$^{3}$, and
\newauthor Steven A. Rodney$^{4}$
\\\\
% List of institutions
$^{1}$School of Physics \& Astronomy, University of Minnesota, 116 Church Street SE, Minneapolis, MN 55455, USA\\
$^{2}$Expertisecentrum voor Digitale Media, Universiteit Hasselt, Wetenschapspark 2, B-3590, Diepenbeek, Belgium\\
$^{3}$Space Telescope Science Institute, 3700 San Martin Drive, Baltimore, MD 21218, USA\\
$^{4}$Department of Physics and Astronomy, University of South Carolina, 712 Main St., Columbia, SC 29208, USA
}

\maketitle

\begin{abstract}
Using the power of gravitational lensing magnification by massive galaxy clusters, the Hubble Frontier Fields provide deep views of six patches of the high redshift Universe. The combination of deep Hubble imaging and exceptional lensing strength has revealed the greatest numbers of multiply-imaged galaxies available to constrain models of cluster mass distributions. However, even with $\mathcal{O}(100)$ images per cluster, the uncertainties associated with the reconstructions are not negligible.  The goal of this paper is to show the diversity of model magnification predictions. We examine 7 and 9 mass models of Abell 2744 and MACS J0416, respectively, submitted to the Mikulski Archive for Space Telescopes for public distribution in September 2015. The dispersion between model predictions increases from 30\% at common low magnifications ($\mu\!\sim\!2$) to 70\% at rare high magnifications ($\mu\!\sim\!40$).  MACS J0416 exhibits smaller dispersions than Abell 2744 for $2\!<\!\mu\!<\!10$. We show that magnification maps based on different lens inversion techniques typically differ from each other by more than their quoted statistical errors. This suggests that some models underestimate the true uncertainties, which are primarily due to various lensing degeneracies. Though the exact mass sheet degeneracy is broken, its generalized counterpart is not broken at least in Abell 2744. Other, local degeneracies are also present in both clusters. Our comparison of models is complementary to the comparison of reconstructions of known synthetic mass distributions. By focusing on observed clusters, we can identify those that are best constrained, and therefore provide the clearest view of the distant Universe.
\end{abstract}


%%%%%%%%%%%%%%%%%%%%%%%%%%%%%%%%%%%%%%%%%%%%%%%%%%

%%%%%%%%%%%%%%%%% BODY OF PAPER %%%%%%%%%%%%%%%%%%

\section{Introduction}

The idea of using galaxy clusters as nature's telescopes goes at least as far back as the 1990's \citep{ham90,bla90}. Since then, numerous works have utilized the magnifying power of clusters to better resolve high redshift galaxies \citep[e.g.][]{hai09,swi07,pet00}. Following in the footsteps of CLASH (Cluster Lensing And Supernova survey with Hubble; PI: M. Postman), Hubble Frontier Fields Survey (HFF; PI: J. Lotz) is an ambitious effort to use massive merging clusters at intermediate redshifts as nature's most powerful telescopes. Launched in 2013, it is a three year project that devotes 840 orbits of Director's discretionary time to do deep imaging of six galaxy clusters plus the accompanying parallel fields. Each field is observed in three HST optical bands and four IR bands. It is the largest commitment to date of HST resources to the exploration of the distant Universe through the power of gravitational lensing.

The HFF project has been remarkably prolific, with contributions in the structure of galaxy clusters \citep{jau14,ric14,ogr15,jau15}, individual high redshift galaxies \citep{mcl15,kaw15,lap16}, and high redshift galaxy luminosity function and its evolution \citep{fin15,coe15,ate15}\footnote{See {\tt http://www.stsci.edu/hst/campaigns/frontier-fields/Publications} for a more comprehensive list of references.}.

To utilize clusters as telescopes one needs to first characterize their uneven optics, i.e. obtain cluster magnification maps. These are derived from the mass distribution maps, which are, therefore, of prime importance to the success of the project. Even with HST's deep imaging of the Frontier Fields, the number of detected strongly lensed sources is not sufficient to map out a cluster's mass distribution with arbitrarily high precision and resolution. A finite observational data set leaves information gaps. How to fill these gaps has been the subject of many papers, based on several lens mass inversion methods. The assumptions going into the different methods are varied (see Section~\ref{mim}); for example, some methods rely on the observed light distribution of cluster galaxies to model the mass of the galaxies and estimate the mass distribution of the cluster, while others do not use cluster galaxies at all.  

Each method generates statistical uncertainties on the recovered mass maps. A practical estimate of the systematic uncertainties is provided by the dispersion between different models. This is the strategy implicitly adopted by HFF. Originally, STScI sponsored five modeling groups with a diverse range of models. These groups have since been joined by others, leading to an even wider range of available models.

In addition to the HFF data products and the published work based on these by various groups, added value of the HFF project comes from two new features that are being adopted for the first time in the cluster lensing community.

{\it (1) Consistency and uniformity}.~
In the past, comparison of various science results from cluster lensing was hindered by the use of different data sets, and differences between model versions from the same groups. Most of that has been eliminated in the HFF project, because teams collectively agree on the most trustworthy set of data to use, namely multiple images, redshifts and weak lensing measurements.

{\it (2) Testing against known mass distributions}.~
Testing of the modeling techniques against synthetic clusters is very important to understand the strengths and weaknesses of the various methods. The synthetic cluster comparison project from \cite{men16} is the largest coordinated effort to date to compare cluster lensing inversion methods based on known mass distributions.  

 In this paper we add a third item to the list: analyses to assess the similarities and differences of the lens inversion methods, without the use of synthetic mass distributions. We hope it will help provide a unified and transparent view of the lens inversion results.

{\it (3) Comparison of models for unknown mass distributions.}~
An exercise parallel to {\it (2)} is to compare reconstructions of an unknown mass distribution, i.e. those of observed galaxy clusters, among themselves. This will show which methods are similar to each other, and which are outliers, and assess which methods are better at estimating systematic uncertainties in the reconstructions, assuming the true uncertainties are given by the scatter among models. This assumption is reasonable because factors other than modeling methods are approximately the same among all the reconstructions. The positions and redshifts of images are the same in all reconstructions, as these were agreed upon by all participating groups. Though the set of sources/images used by various groups are not the same, the number of sources/images used are comparable, see Table~\ref{SLdata}. The quality of the mass reconstructions, as summarized by the lens plane rms (the rms of the differences between the observed and reconstructed image positions) is also similar in all models, and is $<1"$ for most models (see Table~\ref{SLdata}). 

This type of comparison can also provide an indirect test of the state of the art of synthetic models themselves.  If the synthetic clusters are well matched to observed HFF clusters in important respects, then a comparison of reconstructions of observed clusters and that of synthetic clusters should yield statistically similar results.

\begin{figure}    %% figure 1
\centering
\includegraphics[width=0.7\linewidth]{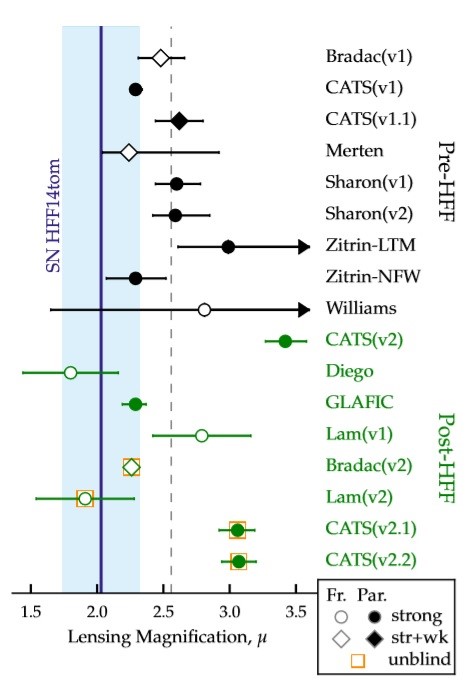}
\vspace{+5pt}
\caption[Observed magnification of SN HFF14Tom]{The observed magnification, $\mu$, of the supernova HFF14Tom (blue highlight) compared to the estimates of various models. Reproduced from \cite{Rodney}, by permission of the AAS. The vertical dashed line marks the unweighted mean for all 17 models, at $\mu=2.6$. The horizontal line segments mark median magnification and 68\% confidence limits of the 17 models.}
\label{fig:Figure1}
\end{figure}

Our current work was inspired by a limited comparison of models presented in \cite{Rodney}, which was based on a supernova Type Ia, HFF14Tom (SN~Tomas), that went off in 2014 behind Abell 2744, one of the HFF clusters. As standard candles, Type Ia supernovae are important because they are the only sources whose actual magnification can be accurately measured, and used to constrain lens inversions~\citep{rie11}. Based on its redshift and light curve properties, \cite{Rodney} deduced that the source was magnified by $\mu=2.03\pm 0.29$ relative to its expected unmagnified brightness. The uncertainty in the measured magnification includes both the photometric uncertainty and the known scatter in the luminosity of Type Ia SNe at that redshift (after correcting for light curve shape and color). Fig.~\ref{fig:Figure1}, taken from that paper, shows the magnification predictions of various lens models for the location and redshift of HFF14Tom, with 68\% uncertainties. The figure also shows that the median of the lensing model magnification predictions is 25\% higher than what is deduced from the observed brightness of the supernova. HFF14Tom is not the first supernova to be detected behind galaxy clusters \citep{ama11,nor14,pat14}, but it is the first to be used systematically as a precise benchmark to compare several models. 

Later in 2014, the Universe provided another opportunity for directly testing lens models with the appearance of the first multiply imaged SN behind the HFF cluster MACS J1149: SN Refsdal~\citep{kel15}. This SN was strongly lensed both by the cluster and also by a single cluster member galaxy, resulting in four distinct images in an "Einstein Cross" configuration. Measurement of the time delays and relative magnifications between these first four images allowed a similar test of lens model accuracy~\citep{rod16,tre16}. The reappearance of SN Refsdal with a fifth image more widely spaced from the initial four is now offering another opportunity to confront lens model predictions with a pencil-beam test~\citep{kel16}. Although these SN tests have been exciting and informative for the lens modeling community, we still have only a handful of examples, scattered across several clusters. HST remains the only observatory that has demonstrated the capability to discover and study these distant SNe in sufficient detail to enable such tests. For the foreseeable future such spot checks will remain few and far between.

In this paper we extend the model comparison analysis of \cite{Rodney} to the whole area of clusters, and especially to regions of high magnification. We also augment the analysis with a range of different statistics. Because the true magnification (the equivalent of the vertical solid line and blue band in Fig.~\ref{fig:Figure1}) as a function of sky location is unknown, we use the median of all models as the comparison benchmark (the equivalent of the vertical dashed line.) We examine all the reconstruction models that used HFF data and were submitted to STScI in September 2015 in response to that call for models.\footnote{Other v3 models were submitted in the months following the deadline, but we do not include these in the current work.} Models v3 and v3.1 of clusters Abell 2744 ($z=0.308$) and MACS J0416 ($z=0.396$) fall into that category.

\begin{table}
\centering
%\begin{threeparttable}[b]
\caption{Summary of input strong lensing data: number of images used by each model, the assumed positional error of the images, and the lens plane rms of the reconstructed maps. The values on either side of the vertical bar are for Abell 2744 and MACS J0416, respectively. All models are v3 unless otherwise noted. (Note that in some models, like GLAFIC, the value of the assumed positional error affects the reconstructions, but in many others, like Sharon/Johnson and Diego, it does not, because the same positional error is assumed for all the images. The assumed positional error may affect the quoted errors in mass reconstructions.)}\label{SLdata}
\begin{tabular}{|c|c|c|c|}
        \hline   Model    & \# images        & assumed   & lens plane \\
                          &  used            & pos. error ($^{\prime\prime}$) & rms ($^{\prime\prime}$)\\
        \hline   CATS     &           52|80  &  0.5|0.5  & 0.50|0.71 \\
        \hline   CATS v3.1&          112|139 &  0.5|0.5  & 0.70|0.54 \\
        \hline   Sharon/Johnson &     81|81  &  0.3|0.3  & 0.42|0.36 \\
        \hline   Zitrin-NFW &         83|93  &  0.5|0.5  & 2.00|2.00$^1$ \\
        \hline   Zitrin-LTM-Gauss &   83|93  &  0.5|0.5  & 1.92|1.99$^1$ \\
        \hline   GLAFIC     &        111|182 &  0.4|0.4  & 0.37|0.44 \\
        \hline   Williams   &         55|88  &  not appl.& 0.29|0.36\\
        \hline   Williams v3.1 & ---\,\,|147 &  not appl.&   ---\,\,|0.27\\
        \hline   Diego     &          16|52  &  0.1|0.1  &  3|3$^2$ \\
        \hline
\end{tabular}
\begin{tablenotes}
\item[1]$^1$The value is somewhat artificially boosted by a lower grid resolution.
\item[2]$^2$Not available, but is typically $3^{\prime\prime}$.
\end{tablenotes}
%\end{threeparttable}
\end{table}

\section{Brief description of lensing mass inversion methods}\label{mim}

In this section we briefly describe each method, highlighting the different model assumptions. These descriptions come primarily from the published source papers, and the {\tt readme} files provided by the modelers and available through the HST Frontier Fields Lensing Models Web-page\footnote{\tt https://archive.stsci.edu/prepds/frontier/lensmodels/}. 

Most lens inversion methods can be classified as parametric, free-form, or hybrid. In a parametric model, the mass distribution is usually tied to the observed light distribution of the cluster galaxy members. Each galaxy's mass distribution is represented by a simply parametrized form, like a pseudo isothermal elliptical mass distribution, PIEMD~\citep{kk93}. The dark matter component is represented by one or two such forms. The total number of model parameters in the methods is usually comparable to the number of lensing constraints. Free-form methods, sometimes called non-parametric to distinguish them from parametric methods, can employ a similar number of parameters, or a vastly greater number. In the latter case, lensing data do not uniquely constrain a lens; instead many realizations are produced, and an average is taken as the solution. The name `free-form' signifies that these methods do not tie the mass distribution of the lens to that of the observed light. Hybrid methods combine features of both approaches. 

 The HFF v3 series of reconstructions use lensed images that were evaluated by all the lens model teams, and collectively assigned a quality, or reliability measure, as GOLD, SILVER or BRONZE. GOLD systems have spectroscopic redshifts and secure multiple image identifications according to most modelers. SILVER systems have no spectroscopic redshifts, but were unanimously classified as secure. BRONZE were considered to be secure systems by most modelers.  All models used the bulk of the GOLD and SILVER images, and some included BRONZE images. (See Table~\ref{SLdata} for the number of sources and images used in each model.) The available astrometric and spectroscopic data for the two clusters is the result of cumulative work by several groups, and is presented in \cite{zit13, jau14, ric14, john14, die15, wan15, rod16}. All modelers used the concordance cosmological model with $H_0=70$ km s$^{-1}$ Mpc$^{-1}$, $\Omega_\Lambda=0.7$ and $\Omega_m= 0.3$.

\vskip0.08in
\noindent{\bf{\small{CATS}}}

The CATS collaboration uses \lenstool, a code that combines strong- and weak-lensing data; reconstructions used here are based on strong lensing only, which uses parametric modeling. The method assumes that the total mass distribution of clusters consists of several smooth, large scale potentials that are modeled by a parametric form, along with contributions from many, typically $N>50$, individual cluster galaxies, modeled using physically motivated parametric forms \citep{jk09}.  Scaling relations are used to tie galaxies' properties to those of $L_\star$ \citep{sch76} galaxies. The models are adjusted in a Bayesian way, probing the posterior probability density with a Markov Chain Monte Carlo (MCMC) sampler \citep{j07}. This process yields error estimates on derived quantities such as the amplification maps and the mass maps~\citep{ric14}. The difference between the two versions of CATS models is that v3 uses the multiple images ranked as GOLD and SILVER by all the lensing teams, while v3.1 uses GOLD, SILVER and BRONZE. See \cite{jau14,jau15} for further details of the reconstructions.

\vskip0.08in
\noindent{\bf{Sharon/Johnson}}  

This team uses a parametric modeling technique, \lenstool~ v6.5 and v6.8 \citep{j07}, which relies on an MCMC to find the best-fit model parameters weighted by Bayesian evidence. The modeling is iterative; it starts by placing cluster-scale masses near the centre of the light distribution, and then builds up the model in complexity by adding more image constraints and more mass components. The cluster and galaxies are modeled as a combination of PIEMD halos. Cluster galaxies are selected by their colors in a color-magnitude diagram, and their parameters are scaled to their luminosity. Galaxies in close proximity to lensed features were modeled separately. The process continues until all the image constraints have been included, and the model root-mean-square residuals (rms) can no longer improve significantly by adding another halo. The early iterations are optimized in the source plane, while the final model is optimized in the image plane, where the rms scatter is computed for each image by tracing through the lens back to the source plane and back out to the image plane \citep[see][for details]{john14}.

\vskip0.08in
\noindent{\bf{Zitrin}} 

The light-traces-mass (LTM) method was first introduced by \cite{br05} and later revised and simplified by \cite{zit09}, where full details can be found. The LTM assumption is adopted for both the galaxies and cluster halo, which are the two main components of the mass model. To start, cluster members are identified from the red-sequence. Each cluster member is then assigned a PIEMD scaled by the galaxy luminosity, so that the superposition of all galaxy contributions constitutes the first component for the model \citep{j07}. This mass map is then smoothed with a Gaussian kernel to obtain a smooth component representing the dark matter density distribution. The two mass components are then added with a relative weight, and supplemented by a two-component external shear to allow for additional flexibility and higher ellipticity of the critical curves. This is the LTM-Gauss model. In the NFW model, instead of smoothing the galaxy distribution to obtain the cluster-wide component, the latter is represented with elliptical NFW density profiles, following~\cite{nfw97}.  The best fit solution and accompanying errors are estimated via a long MCMC; see \cite{zit13} for further details.

\begin{figure*}    %% figure 2
\centering
\includegraphics[width=0.20\textwidth]{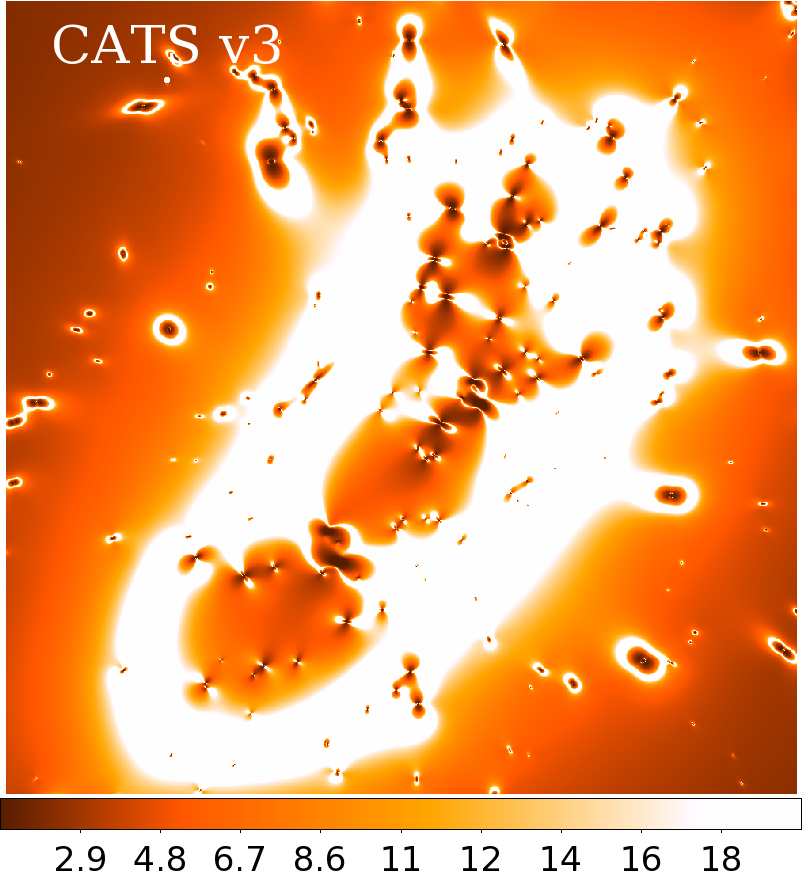}
\vspace{1pt}
\includegraphics[width=0.20\textwidth]{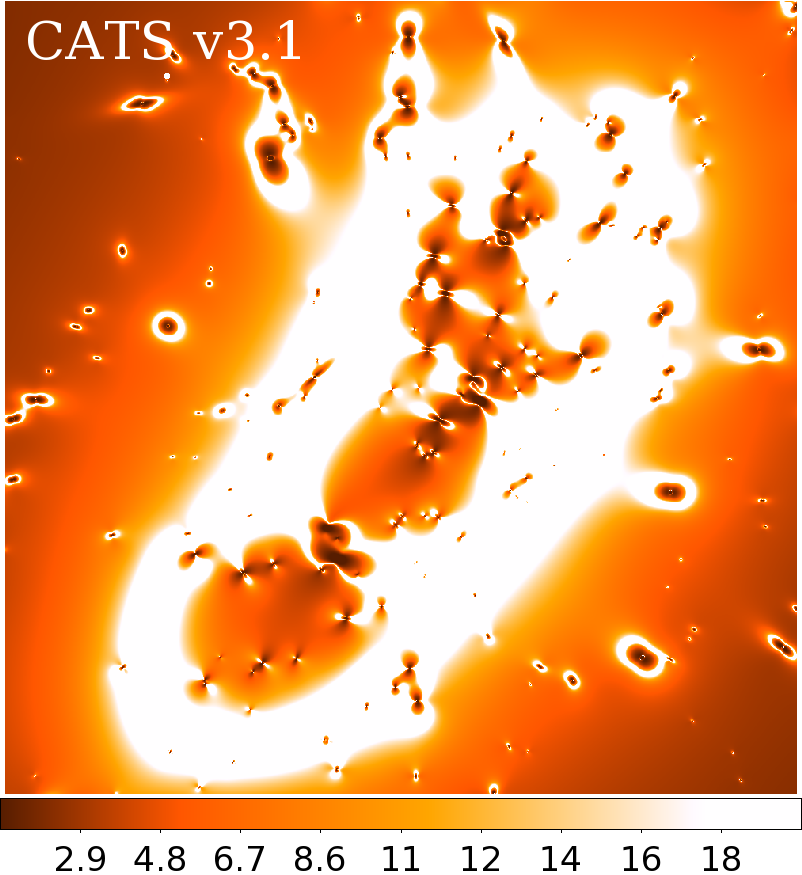}
\vspace{1pt}
\includegraphics[width=0.20\textwidth]{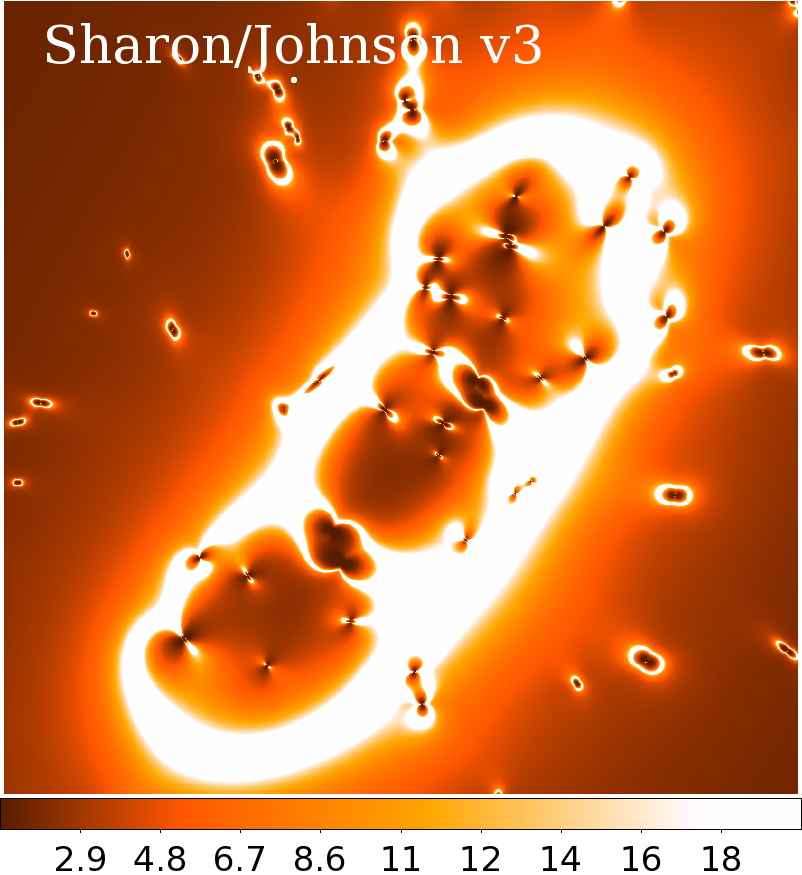}
%\hfill
\includegraphics[width=0.20\textwidth]{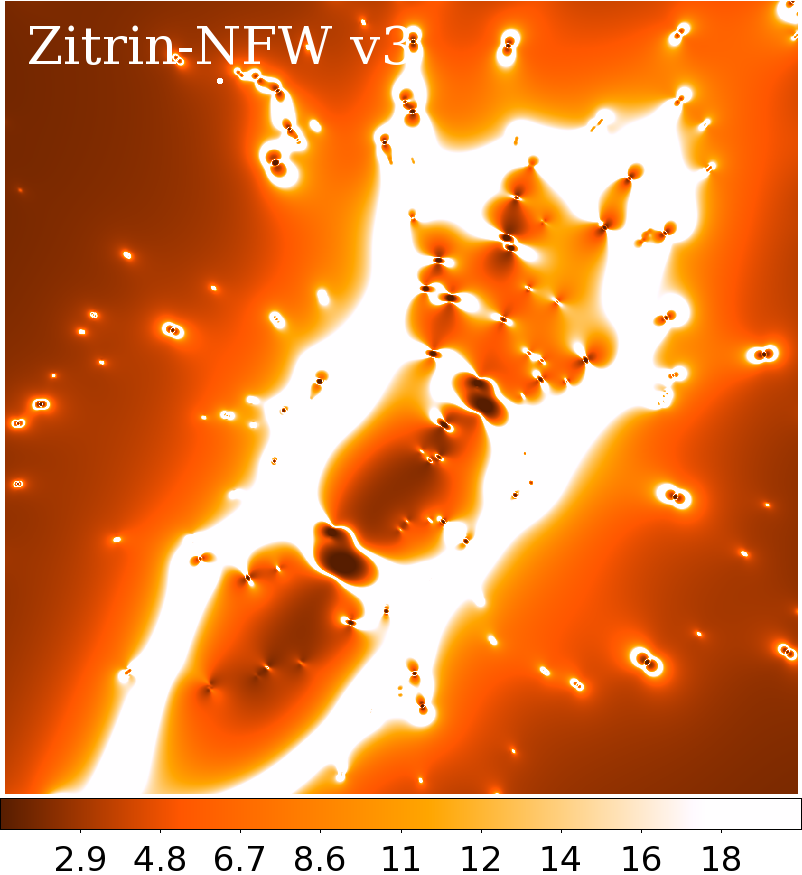}
\vspace{1pt}
\includegraphics[width=0.20\textwidth]{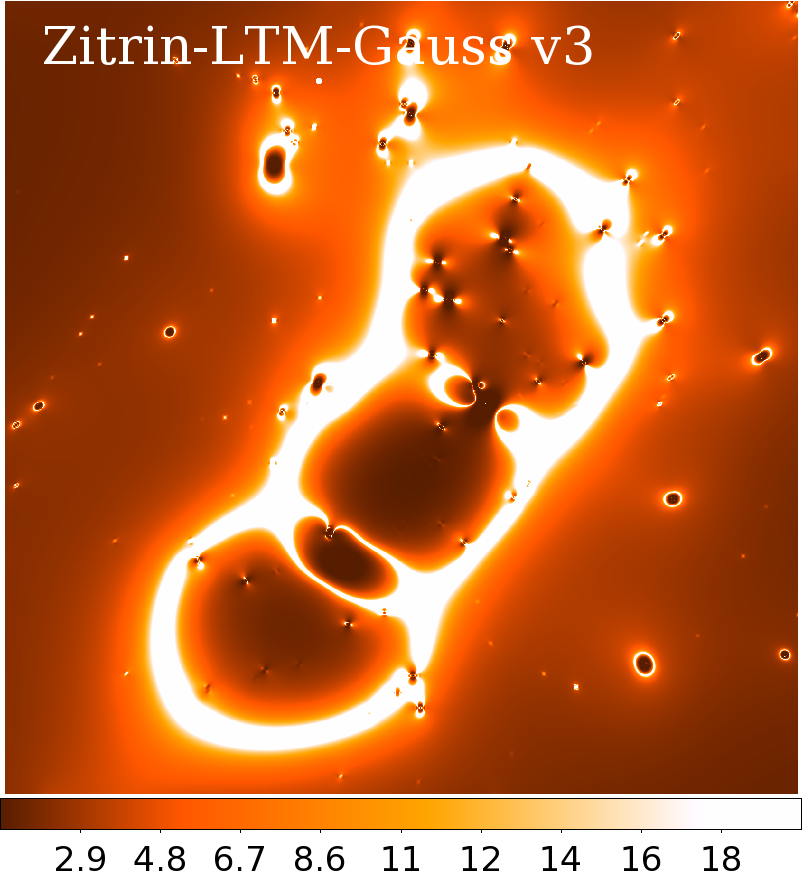}
\vspace{1pt}
\includegraphics[width=0.20\textwidth]{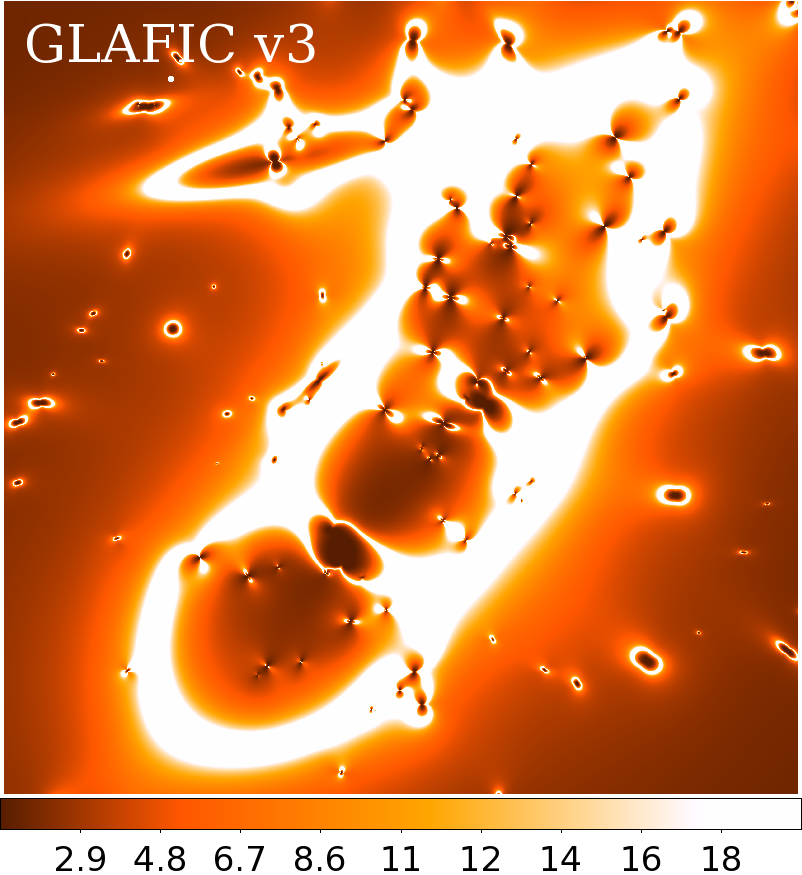}
%\hfill
\includegraphics[width=0.20\textwidth]{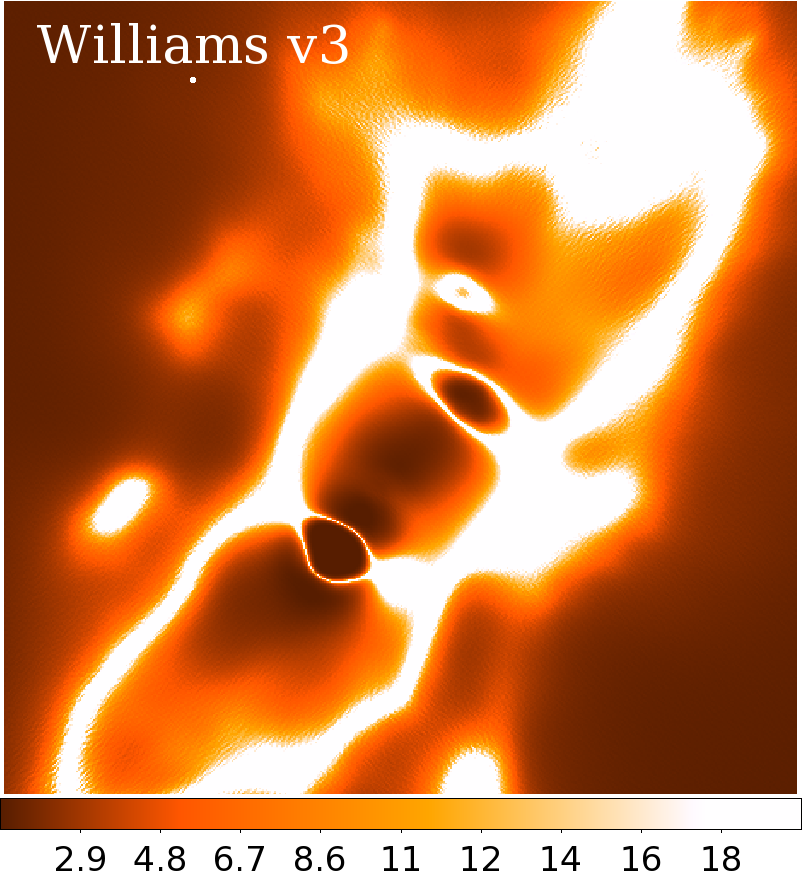}
\vspace{+15pt}
\caption{Magnification distributions of the `best' maps for each of the models of Abell 2744. The color scale is the same in all panels, and goes from magnification of 1 (darkest color) to 20 (white). Each panel is a square $100^{\prime\prime}$ on the side, centred on RA$=3.58966^\circ$, Dec$=-30.39994^\circ$.}
\label{fig:magnifsA2744}
\end{figure*}

\begin{figure*}    %% figure 3
\centering
\includegraphics[width=0.20\textwidth]{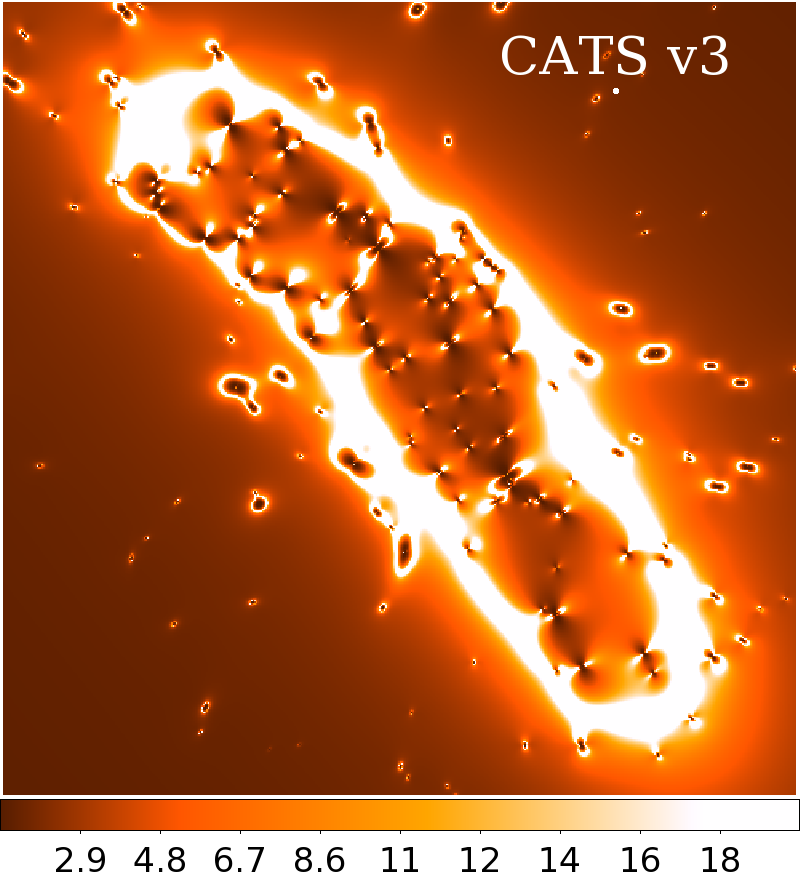}
\vspace{1pt}
\includegraphics[width=0.20\textwidth]{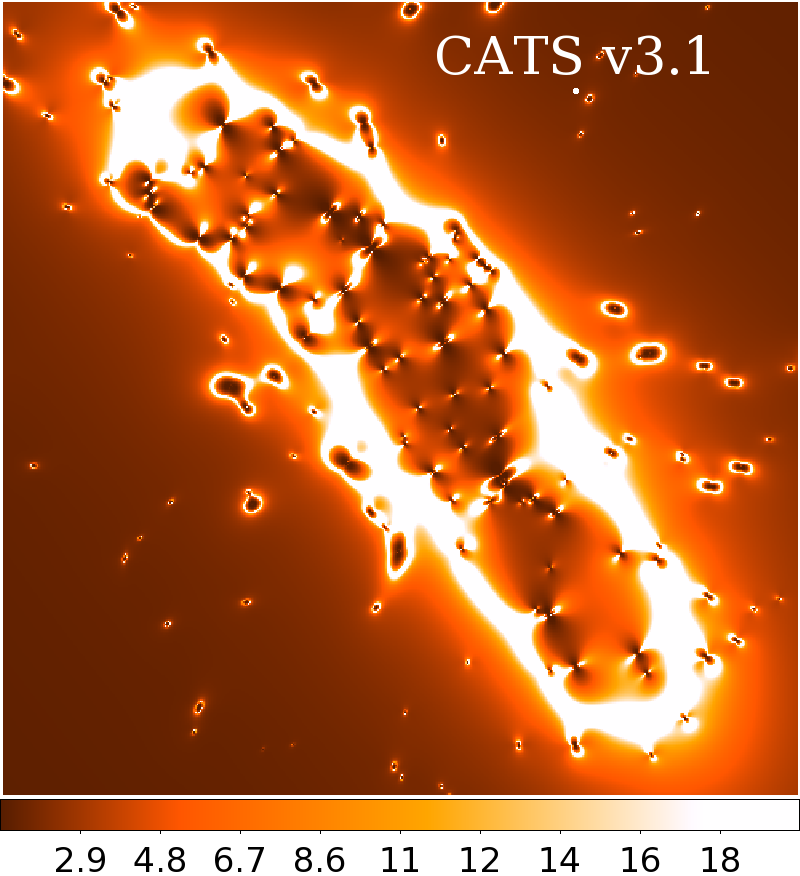}
\vspace{1pt}
\includegraphics[width=0.20\textwidth]{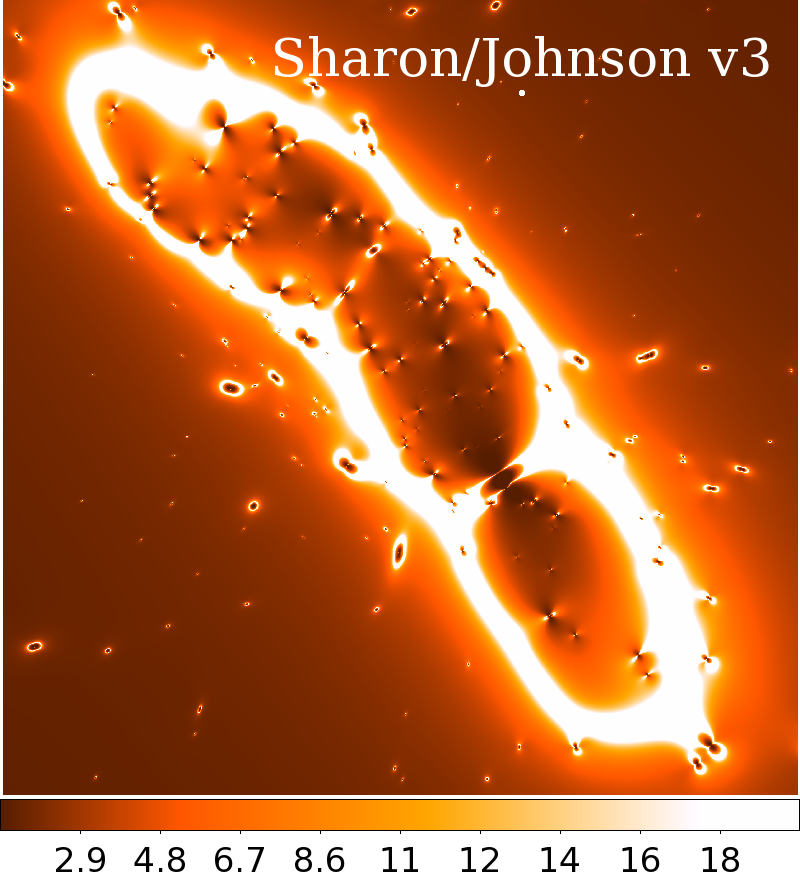}
%\hfill
\includegraphics[width=0.20\textwidth]{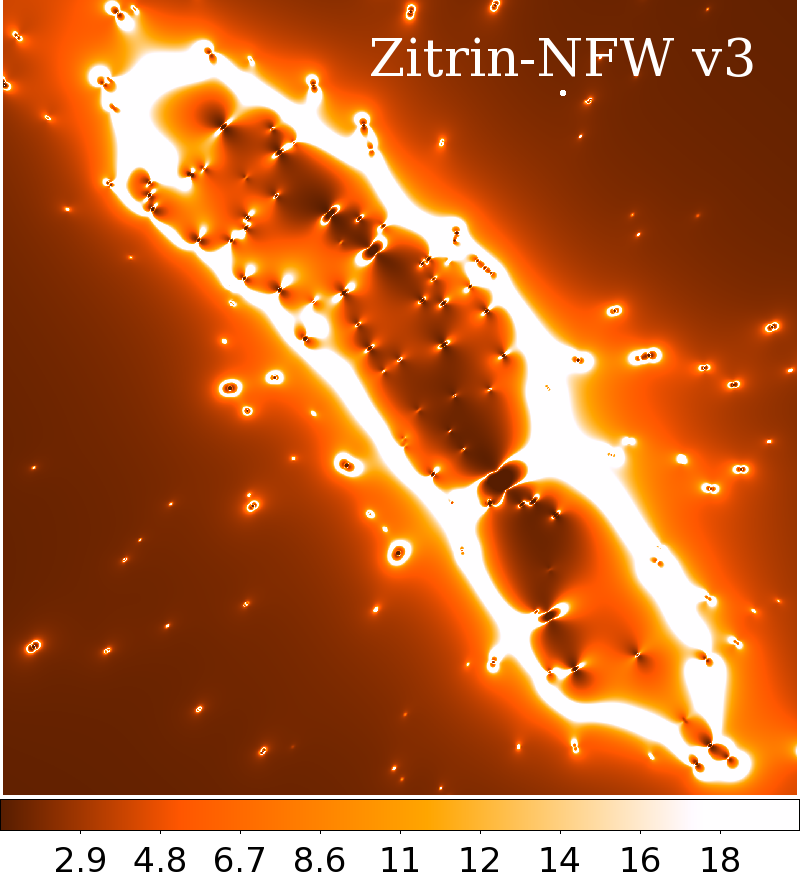}
\vspace{1pt}
\includegraphics[width=0.20\textwidth]{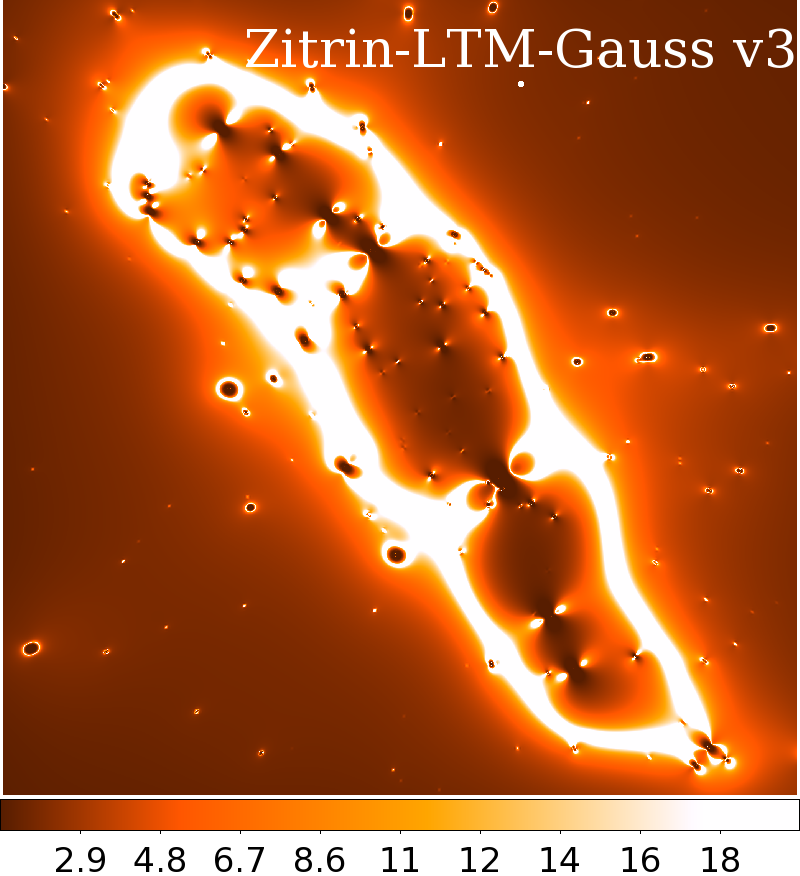}
\vspace{1pt}
\includegraphics[width=0.20\textwidth]{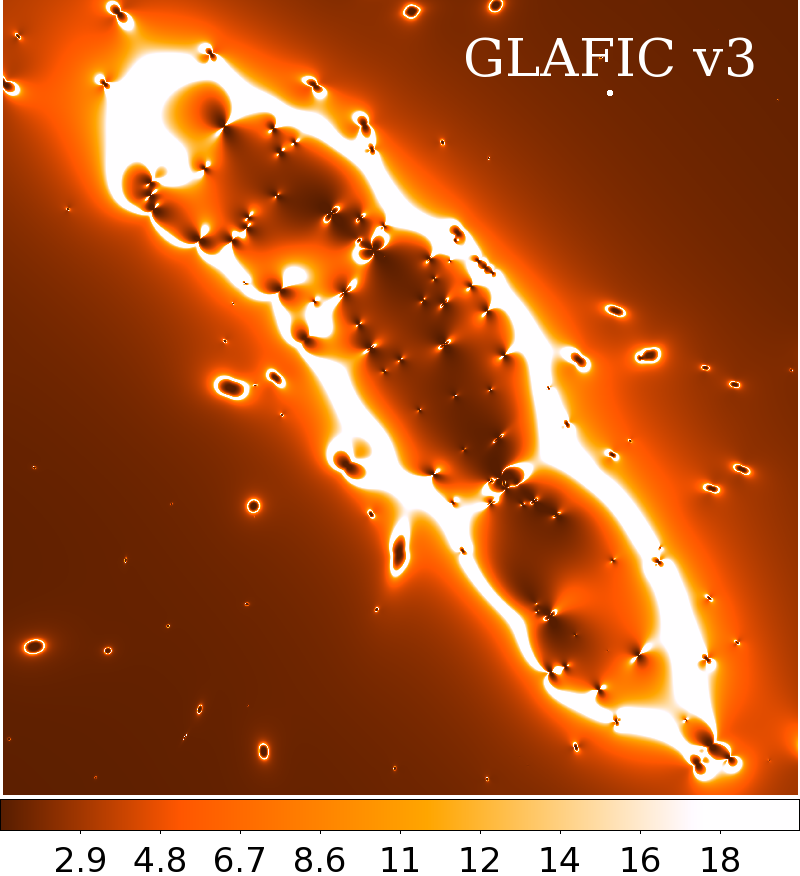}
%\hfill
\includegraphics[width=0.20\textwidth]{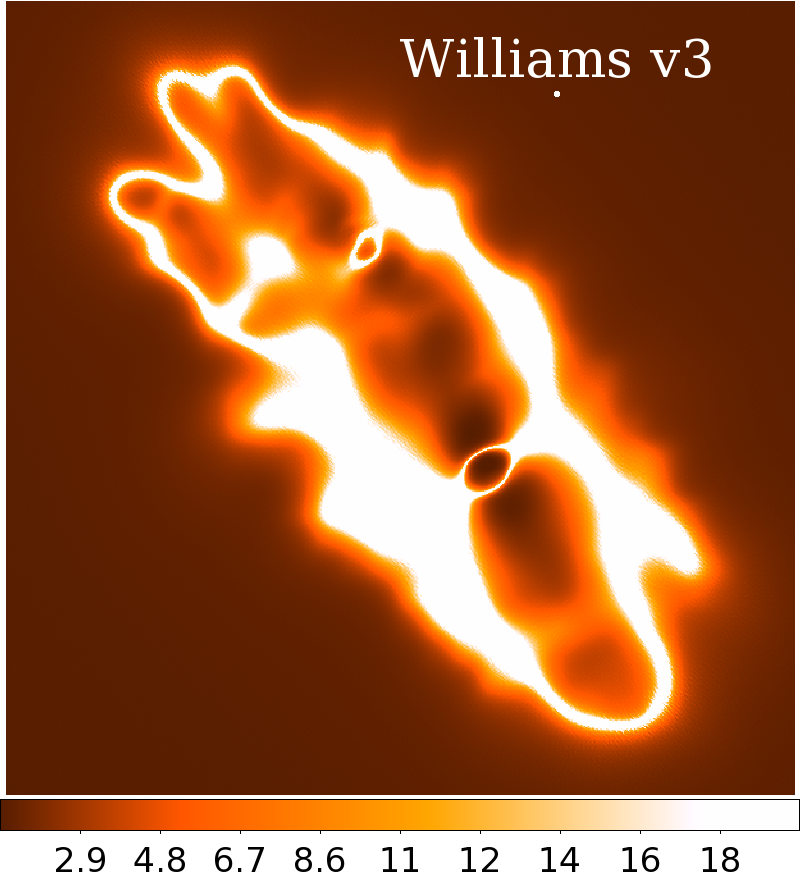}
\vspace{1pt}
\includegraphics[width=0.20\textwidth]{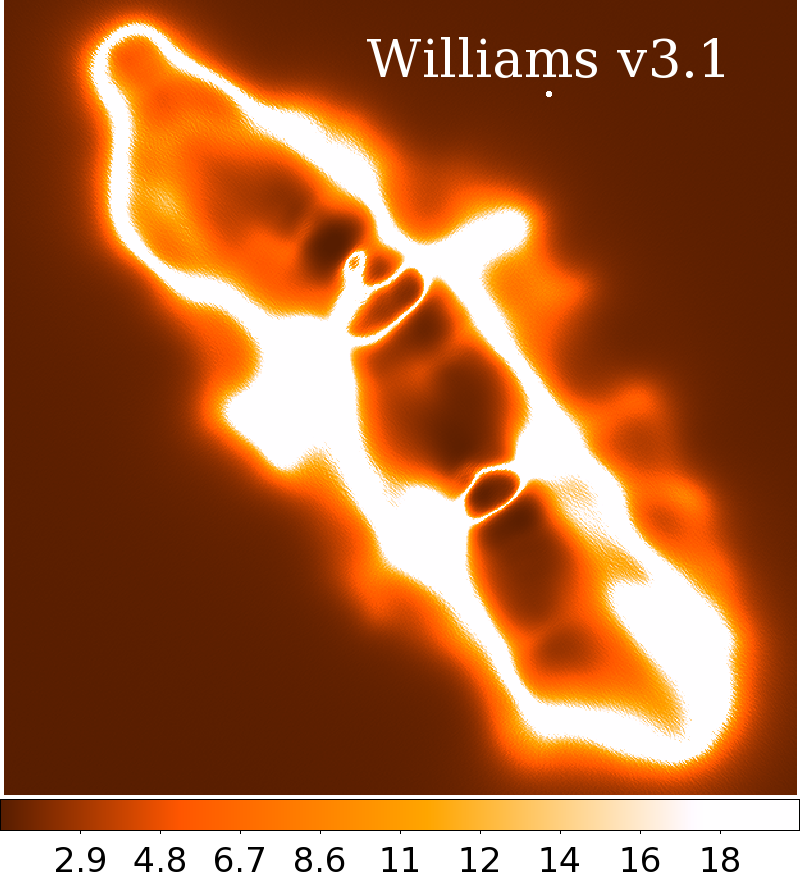}
\vspace{1pt}
\includegraphics[width=0.20\textwidth]{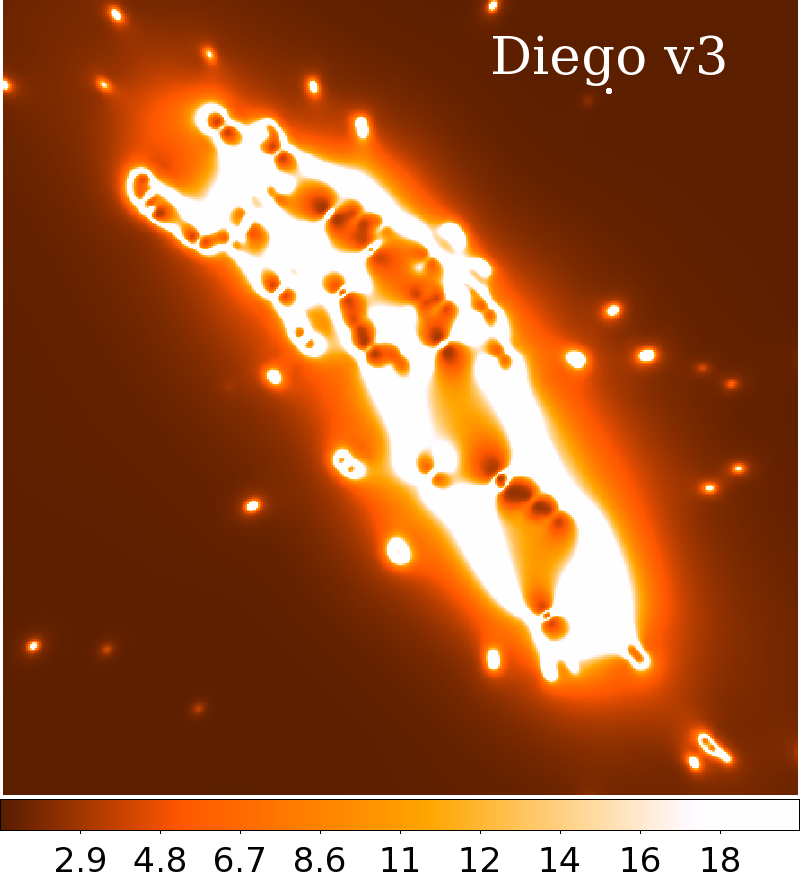}
\vspace{+15pt}
\caption{Same as Fig.~\ref{fig:magnifsA2744}, but for MACS J0416. Each panel is a square $125^{\prime\prime}$ on the side, centred on RA$=64.03680^\circ$, Dec$=-24.07390^\circ$.}
\label{fig:magnifsMACS0416}
\end{figure*}

\vskip0.08in
\noindent{\bf{\small{GLAFIC}}} 

This analysis uses GLAFIC \citep{ogu10}, a parametric modeling technique. The cluster wide components are represented by elliptical NFW density profiles, while the member galaxies are modeled by pseudo-Jaffe ellipsoids\citep{kee01}. To reduce the number of parameters, scaling relations are used to tie galaxies' properties to those of $L_\star$ galaxies. This is common to many methods that use cluster galaxies, for example, ~\lenstool. The properties of some cluster galaxies that are located close to lensed images are not scaled to $L_\star$, but are modeled independently, also using pseudo-Jaffe profiles. The number of these galaxies is a few, at most. In addition to these two components, GLAFIC also uses external and internal perturbations of the lensing potential, both of which are represented by a multipole Taylor expansion of the form, $\phi=(C/m)r^n\,\cos[m(\theta-\theta_\star)]$, at the position of the BCG. A downhill simplex method is used to simultaneously optimize all model parameters in the source plane, and find the best fit models. MCMC is then used to estimate uncertainties. The detailed modeling method and results are described in \cite{kaw16}. 

\vskip0.08in
\noindent{\bf{\small{WSLAP+}}}

The WSLAP+ reconstruction method of \cite{die15} is mostly free-form, but does include mass contribution from individual cluster member galaxies, whose parametrized mass distribution is tied to their light. The non-parametric part of the model represents the large scale mass distribution of the lens, and makes use of the fact that the strong lensing equations are linear in the unknowns, namely, the weight of the mass components, and the positions of the sources. The modeling field of view is divided into a $\sim 32\times 32$ grid, each containing a two-dimensional Gaussian. The weights of these Gaussians, as well as the parameters describing the galaxies and the source positions are found using quadratic programming. The reconstructions used the GOLD and SILVER data sets of images.

\vskip0.08in
\noindent{\bf{Williams/\grale}}

\grale, described in \cite{lie06} and \cite{lie07}, is a free-form lens reconstruction method that uses an adaptive grid and, in contrast to all other methods described here, uses no information about the cluster galaxies. It uses a genetic algorithm to iteratively refine the mass map solution. An initial coarse grid is populated with a basis set, such as projected Plummer density profiles, combined with a uniform mass sheet covering the whole modeling region. The code is started with an initial set of trial solutions. As the code runs the more dense regions are resolved with a finer grid, where each cell is given a Plummer profile with a proportionate width. These solutions, as well as all the later evolved ones, are evaluated for genetic fitness and the fit ones are cloned, combined and mutated. The final map consists of a superposition of a mass sheet and many Plummers, typically several hundred, each with its own size and weight. The dispersion between different \grale~ runs quantifies mass uncertainties, which are due to mass degeneracies present when all image information is held fixed. \grale~ can be used in two modes: treating lensed images as extended (v3), or point-like (v3.1). The reconstructions used most of the GOLD and SILVER lensed images, as well as a few BRONZE.

\begin{figure*}    %% figure 4
\centering
\includegraphics[width=8.7cm]{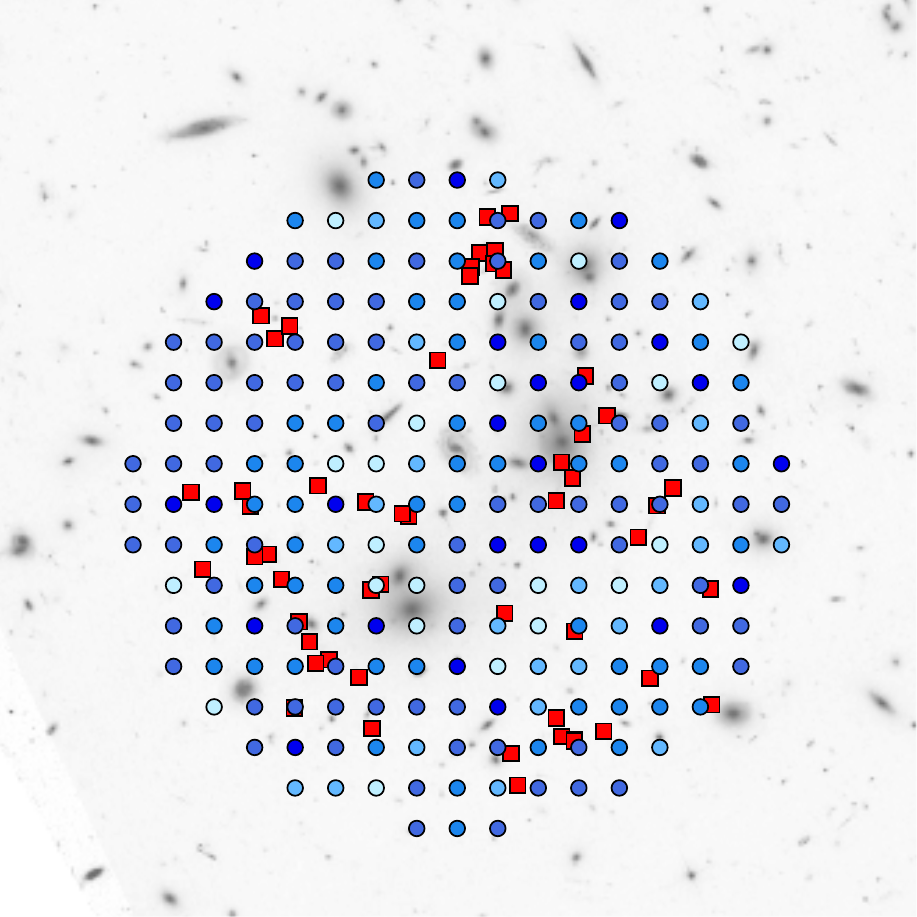}
\hfil
%\hskip0.05cm
\includegraphics[width=8.7cm]{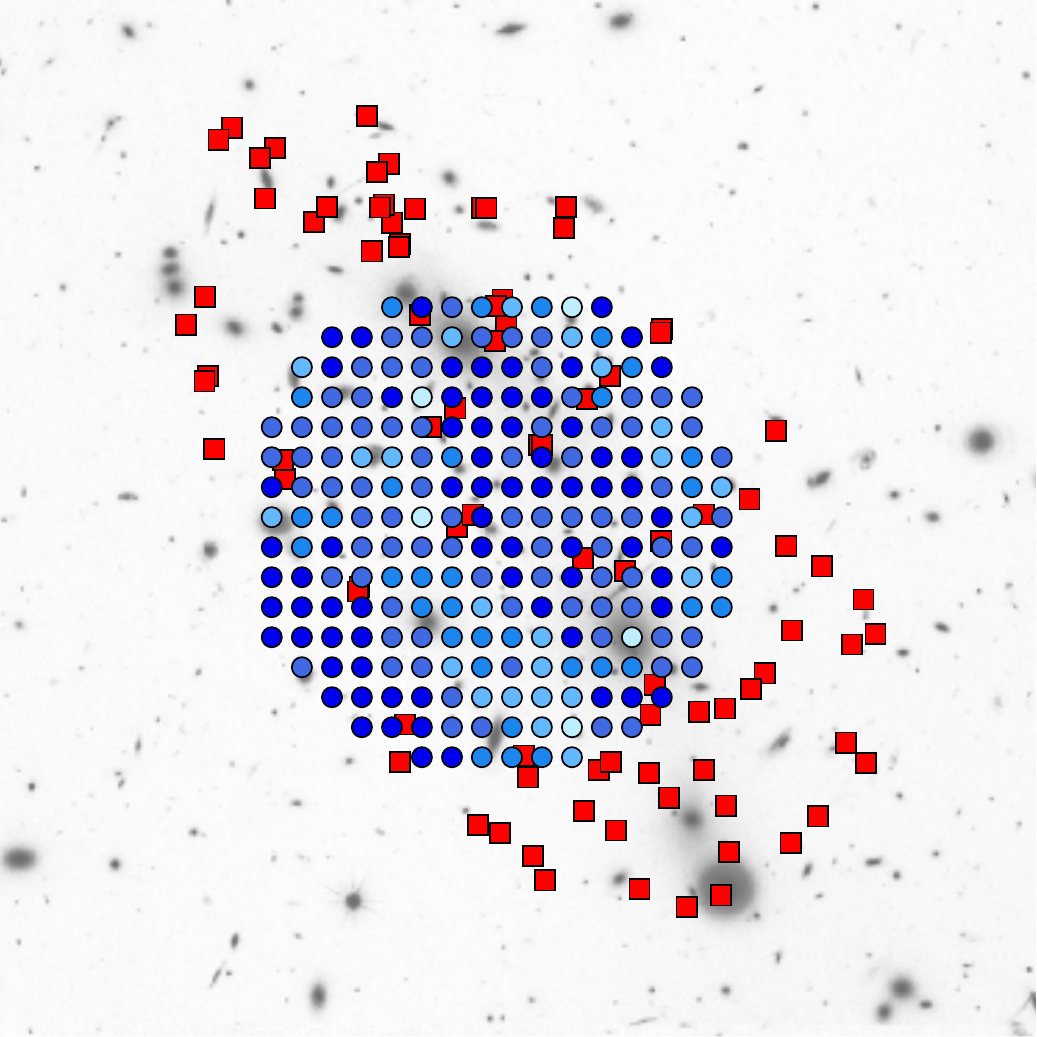}
\caption{The locations of $214$ grid points (shown in blue) where magnifications were extracted. These were used for most of the analysis in this paper. Lensed images of the GOLD and SILVER quality are shown as red squares. The shade of the blue points encodes the fractional Normalized Median Absolute Deviation (fNMAD) at each point, in steps of 25\%: the darkest (lightest) blue has fNMAD$<$25\% ($>$100\%). The median (over all locations) of these values is 31\% and 41\% for Abell 2744 and MACS J0416, respectively; see Section~\ref{comp}. The sizes of the two fields are comparable to those of Figs~\ref{fig:magnifsA2744} and \ref{fig:magnifsMACS0416}, respectively.}
\label{ds9}	
\end{figure*}
 
\section{Comparison of models}\label{comp}

 The goal of the paper is to give the reader a quantitative and qualitative impression of the diversity in magnification predictions obtained by various reconstruction methods. We examine all mass models that were submitted to STScI in the latest round, i.e. v3 and v3.1. Two clusters were part of that call, Abell 2744 and MACS J0416.  Each reconstruction team generated many realizations, for example, from the MCMC chain, for a given cluster, using the same model. For a given model, the Frontier Fields magnification web tool\footnote{The web tool can be found here:\\{\tt https://archive.stsci.edu/prepds/frontier/lensmodels/webtool/}} allows the user to obtain the best\footnote{Each modeling team was asked to provide a range of individual reconstructions, as well as the best reconstruction. How best reconstructions were obtained differs between teams.} and the median magnification estimates for any sky location and source redshift, as well as the uncertainty for any user-specified percentage range. 

As a visual reference, Figs~\ref{fig:magnifsA2744} and \ref{fig:magnifsMACS0416} present magnification maps of the best reconstructions for all the models of both the clusters. Though we do not use `best' maps in the rest of analysis, they provide a convenient at-a-glance comparison of magnification patterns.

To compare models for a given cluster we select a set of sky locations and record each model's median magnification and the accompanying 68\% uncertainty range, for each of these locations. We denote median magnifications for each model and sky location by $\mu_m(\vec\theta)$, where subscript $m$ stands for `model'. Note that most teams presented one model, but some submitted two, v3 and v3.1, so `model' and `team' are not always synonymous. (`Model' denotes a set of 100 or so individual reconstructions for a given cluster, by a given team, using the same inputs and assumptions.) The 68\% uncertainties for a given model are denoted by $\delta_m^+(\vec\theta)$ and $\delta_m^-(\vec\theta)$, or just $\delta_m(\vec\theta)$. All magnifications are for sources at $z=9$.

At each location we find the magnification which represents the median of all models' medians. Because the number of submitted models for Abell 2744 and MACS J0416 are both odd---7 and 9 respectively---the median of all the models is the middle of the ordered list of magnifications, one from each reconstruction model. This median at a given $\vec\theta$ location is called $\tilde\mu(\vec\theta)$. Because the true magnification at each location is unknown, we use $\tilde{\mu}$ as the comparison benchmark, whereas SN tests (e.g. Rodney et al. 2015, 2016) used the actual magnification of the supernova at one location. While $\tilde{\mu}$ need not be the true magnification, it is, arguably, the best guess. Below, we present a few different types of statistical measures to quantify how the models compare to each other.

We concentrate on the central regions of clusters because these are best constrained by the multiple images. Magnifications in these regions are generally $\simgt 2$, and can be as high as several hundred. For Abell 2744 our sky locations are chosen from within a circle centred on the cluster centre (RA$=3.59025^\circ$, Dec$=-30.40224^\circ$) with a radius of 40 arcseconds. For MACS J0416, the radius of the circle is 30 arcseconds, and the cluster center is at RA$=64.03698^\circ$, Dec$=-24.07371^\circ$ (J2000).

Within this circle, the points are spaced on a regular square grid; we use a total of 214 sky locations per cluster. Fig.~\ref{ds9} shows these points in blue, for the two clusters. The shade of blue points encodes the fractional Normalized Median Absolute Deviation (fNMAD) at each point, or NMAD/$\tilde\mu(\vec\theta)$, where
\begin{equation}
{\rm NMAD}=1.4826\times{\rm median}|\mu_{m}(\vec\theta)-\tilde\mu(\vec\theta)|.
\label{eqNMAD}
\end{equation}
The factor 1.4826 is appropriate if one assumes that the model magnification estimates are normally distributed -- which is not necessarily true here.  The five shades of blue are in steps of 25\%: the darkest (lightest) blue has fNMAD$<$25\% ($>$100\%). The median (over all locations) of these values is 31\% and 41\% for Abell 2744 and MACS J0416, respectively.  The GOLD and SILVER lensed images are shown in red. 

\begin{figure*}    %% figure 5
\centering
\includegraphics[width=0.8\textwidth, totalheight=0.4\textheight]{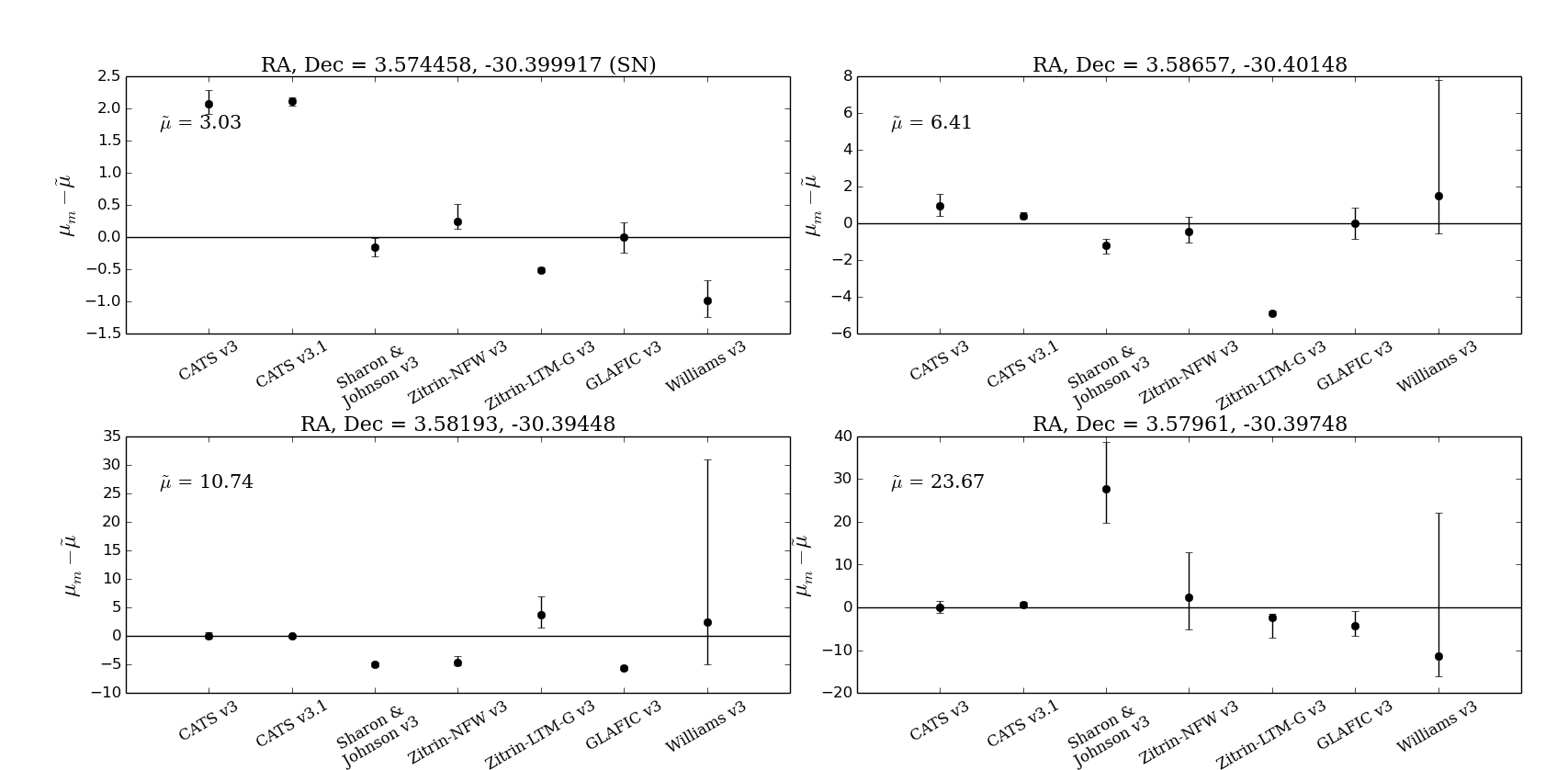}
\caption{Examples of each model's median magnification predictions, $\mu_m$, plotted as deviations from the median of all models, $\tilde{\mu}$, at four sky locations in Abell 2744. (See the beginning of Section~\ref{comp} for the definition of a `model'.) The error-bars represent 68\% confidence limits. The top left panel is for the location of SN HFF14Tom \citep{Rodney}, but not for its redshift; $z_s=9$ is used instead, the same as for all the analyses in this paper. The other three locations were chosen randomly from those marked by the blue points in the left panel of Fig.~\ref{ds9}.}
\label{fig:Figure3}
\end{figure*}

\begin{figure*}    %% figure 6
\centering
\includegraphics[width=0.8\linewidth, totalheight = 0.4\textheight]{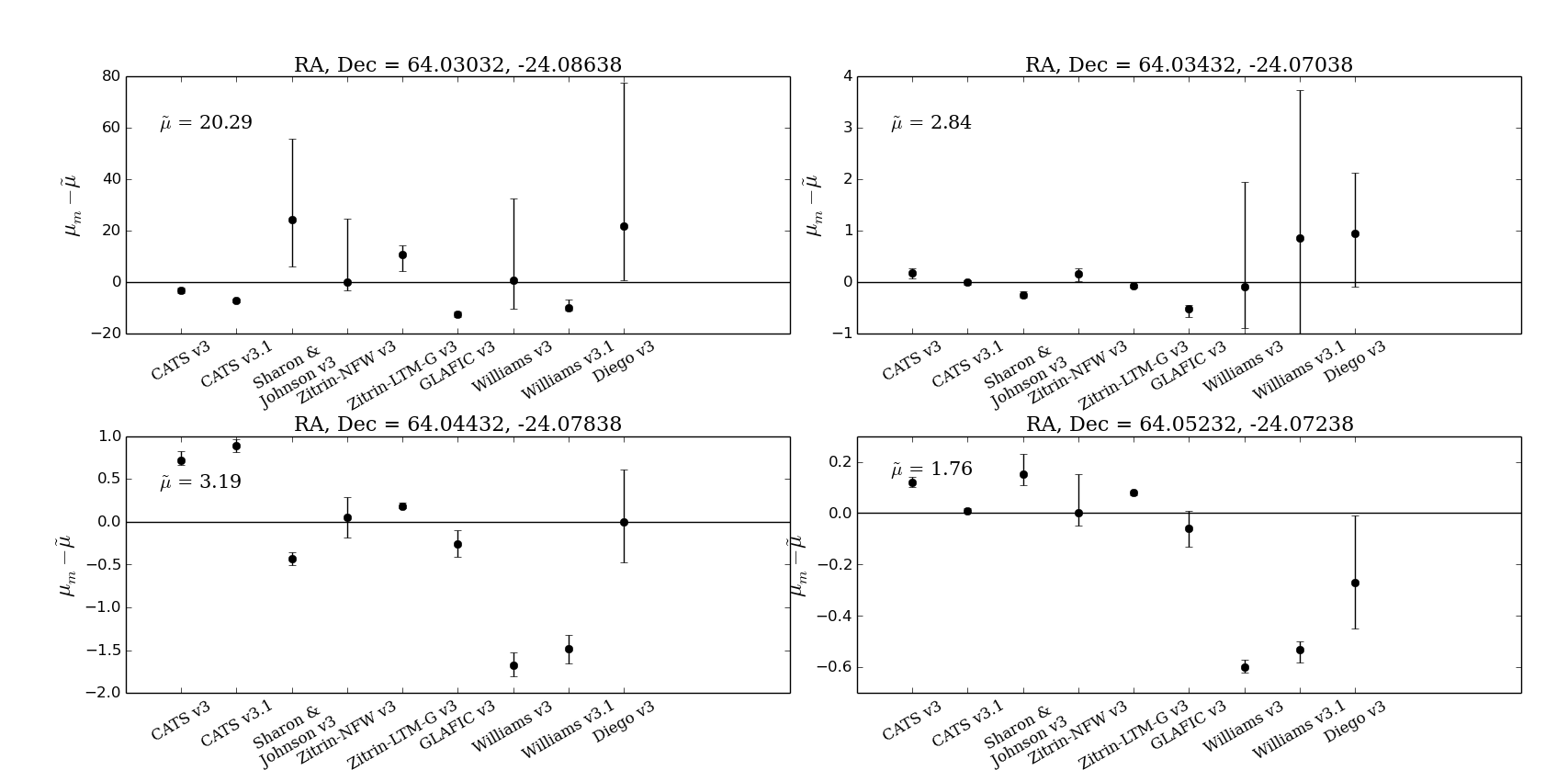}
\caption{As Fig.~\ref{fig:Figure3}, but for MACS J0416. Here, all four sky locations were chosen randomly from those marked by the blue points in the right panel of Fig.~\ref{ds9}.}
\label{fig:Figure4}
\end{figure*}

Figs~\ref{fig:Figure3} and \ref{fig:Figure4} show the same type of plot as fig.~6 in \cite{Rodney}, but for a handful of randomly selected sky locations within Abell 2744 and MACS J0416. The median of all models at that location, $\tilde{\mu}$ is shown in the upper left of each panel, and the vertical axis shows $\mu_m - \tilde{\mu}$. These figures show that there is a considerable amount of scatter in predicted magnifications, but the degree of scatter, and how well it is captured by the error-bars varies between locations.

\begin{figure*}    %%figure 7
	\centering
	\includegraphics[width=1.05\linewidth, totalheight=0.3\textheight]{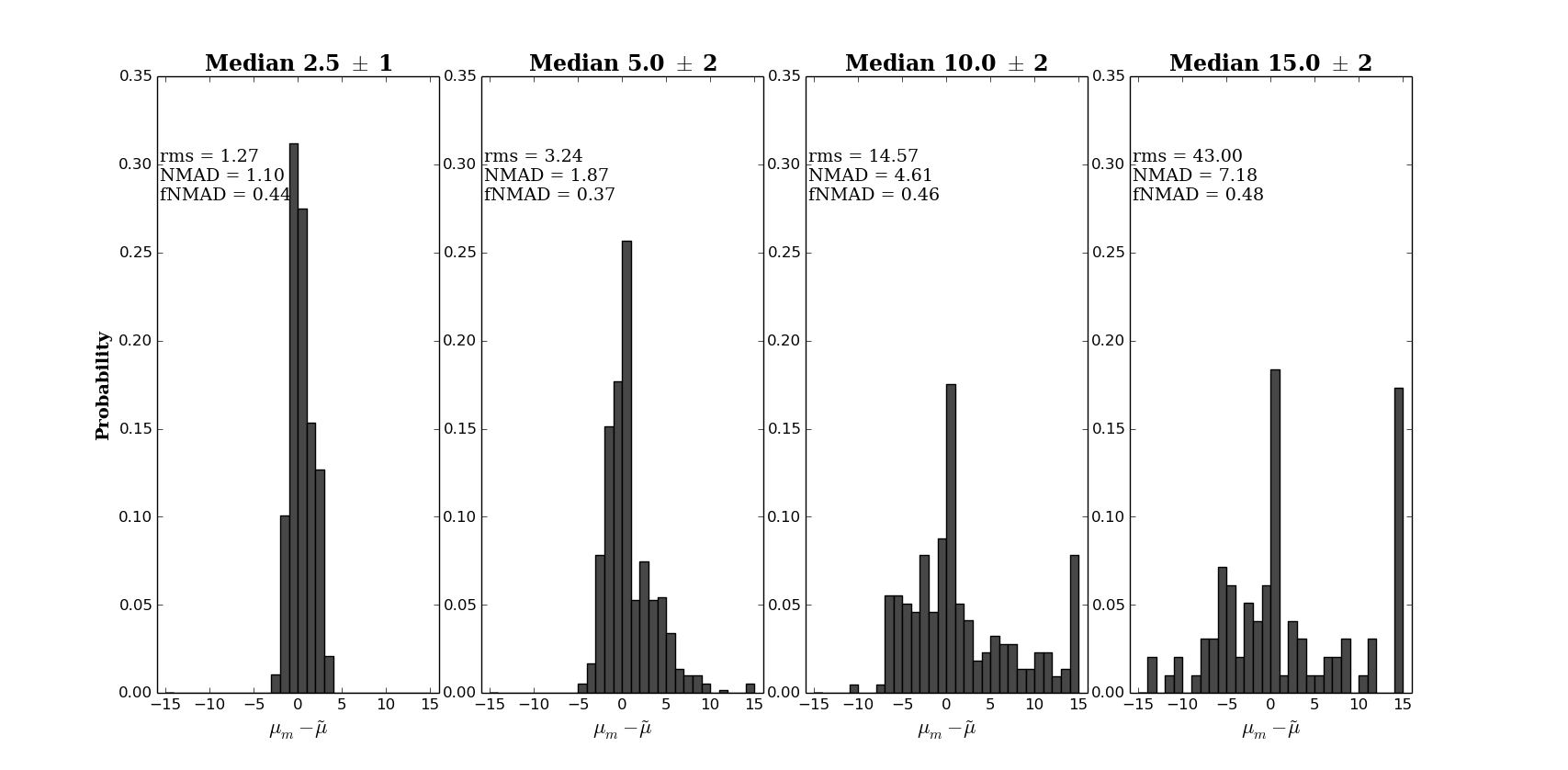}
    \caption{Each of the four panels show $\mu_m-\tilde\mu$ for all 7 models (see the beginning of Section~\ref{comp} for the definition of a `model'), at sky locations within Abell 2744, where $\tilde\mu$ is $2.5\pm1$, $5\pm2$, $10\pm2$ and $15\pm2$, respectively. The sky locations were selected from those marked by the blue points in the left panel of Fig.~\ref{ds9}. The lowest $\tilde\mu$ value is similar to that where HFF14Tom appeared. The histograms are clipped at $\pm 15$, and the values that happen to lie outside of this range are piled up at the edges of the plot boxes (see Section~\ref{comp}). Each panel displays three values quantifying the width of the distribution: the rms, the Normalized Median Absolute Deviation, NMAD, eq.~(\ref{eqNMAD}), where the median is over the sky locations that make up the histogram, and the fractional NMAD, fNMAD = NMAD/$\tilde\mu$.}
\label{fig:Figure11}
\end{figure*}

\begin{figure*}    %% figure 8
	\centering
	\includegraphics[width=1.05\linewidth, totalheight=0.3\textheight]{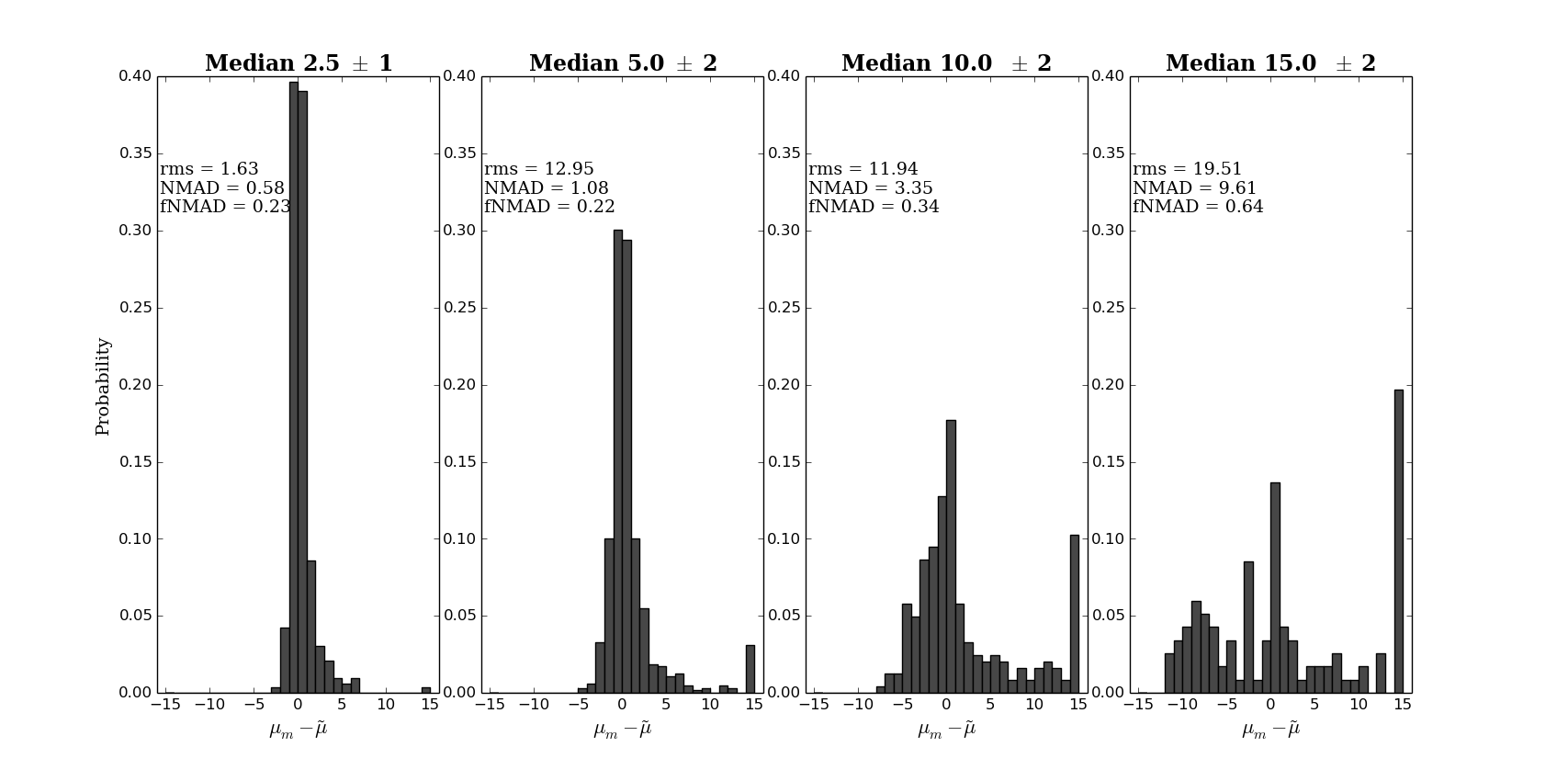}
		\caption{Same as Fig.~\ref{fig:Figure11}, but for all 9 models of MACS J0416.}
		\label{fig:Figure12}
\end{figure*}

An immediate generalization of these plots is to group the data presented in the panels of Figs~\ref{fig:Figure3} and \ref{fig:Figure4} by $\tilde\mu$. One can ask: What is the systematic error in estimated magnification for sky locations where magnification is around some fixed value? Figs~\ref{fig:Figure11} and \ref{fig:Figure12}, for Abell 2744 and MACS J0416 respectively, pick out sky locations where the median of all models, $\tilde\mu$ is $2.5\pm1$, $5\pm2$, $10\pm2$ and $15\pm2$, and plot the distribution of all models' magnifications in the four panels. Some histograms have very large $|\mu_m-\tilde{\mu}|$ values. To account for these without extending the horizontal axes to unreasonable lengths, we pile up all the $|\mu_m-\tilde{\mu}|>15$ values at the corresponding positive or negative edges of the panel boxes. This hides the information about the actual value of $\mu_m-\tilde{\mu}$ (above $15$ or below $-15$), but does represent the total number of these points. There is considerable dispersion in these magnifications, with some distributions extending to magnifications as low as 1, and as high as $2\tilde\mu$. 

\begin{figure}     %% figure 9
\centering
\includegraphics[width=0.95\linewidth]{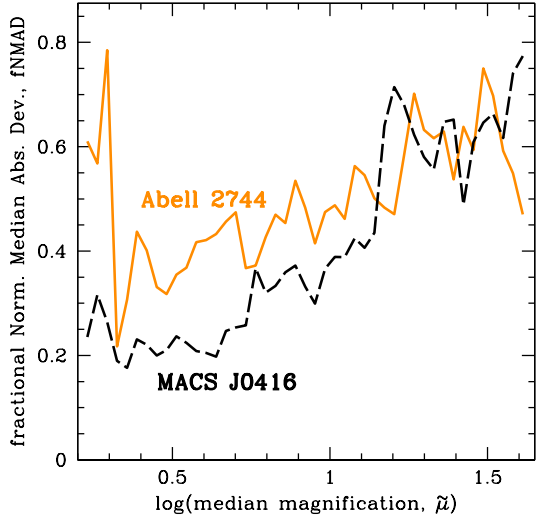}
\vskip-0.7in
\caption{Fractional Normalized Median Absolute Deviation, ~~fNMAD, for sets of sky locations whose median magnification, $\tilde\mu$, is within $\pm$10\% of that value. Because of the logarithmic bin widths, the values plotted here are somewhat different from the ones presented in the four panels of Figs~\ref{fig:Figure11} and \ref{fig:Figure12}.}
\label{smfNMADvsmedian}
\end{figure}

\begin{figure*}     %% figure 10
\includegraphics[width=1.0\linewidth, totalheight=0.4\textheight]{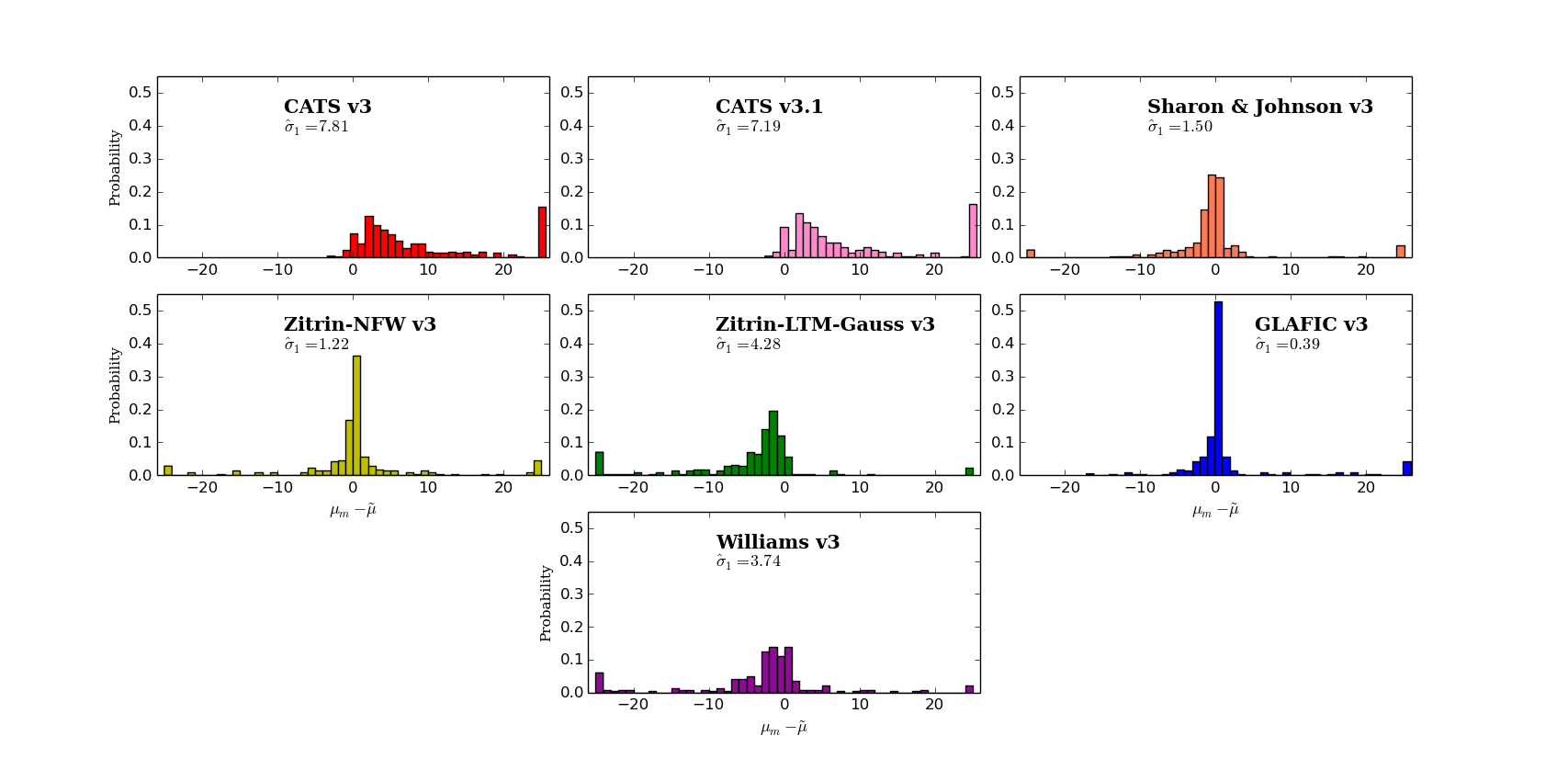}
\caption{The deviations, $\mu_m-\tilde\mu$, of each model's magnification predictions, $\mu_m$, from the median of all models at the corresponding sky locations, $\tilde\mu$, for Abell 2744. Magnification deviations that fall beyond $\pm 25$ are piled up at either end of each panel. Column 2 of Table~\ref{table2} shows the widths of these distributions (also displayed in each panel), quantified by the Normalized Median Absolute Deviation (see Section~\ref{comp}).}\label{fig:Figure5}
\end{figure*}	

\begin{figure*}     %% figure 11
\includegraphics[width=1.0\linewidth, totalheight=0.4\textheight]{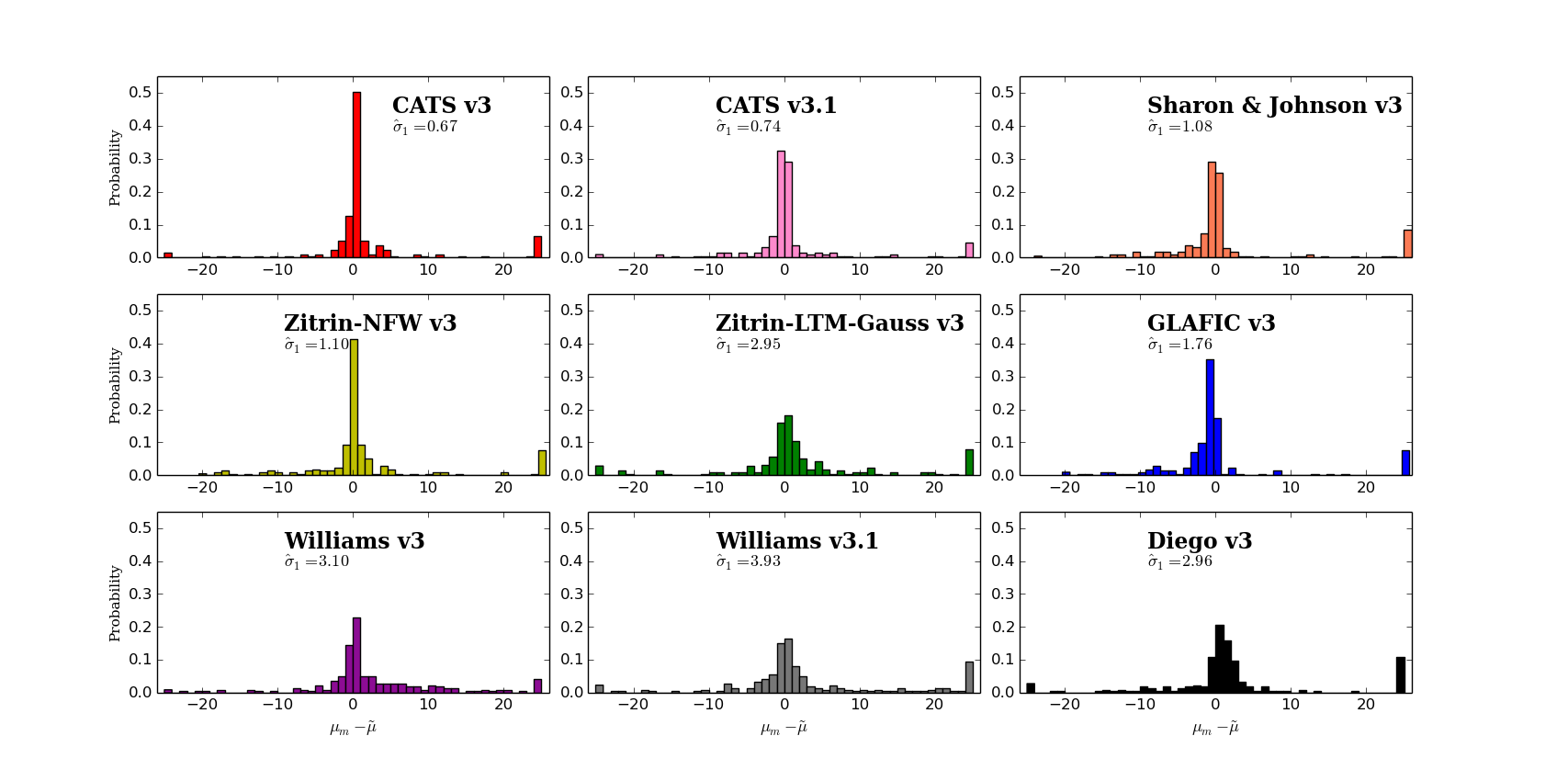}
\caption{Same as Fig.~\ref{fig:Figure5}, but for MACS J0416. Column 4 of Table~\ref{table2} shows the widths of these distributions (also displayed in each panel), quantified by the Normalized Median Absolute Deviation (see Section~\ref{comp}).}\label{fig:Figure6}
\end{figure*}	

\begin{figure*}     %% figure 12
\includegraphics[width=1.0\linewidth, totalheight=0.4\textheight]{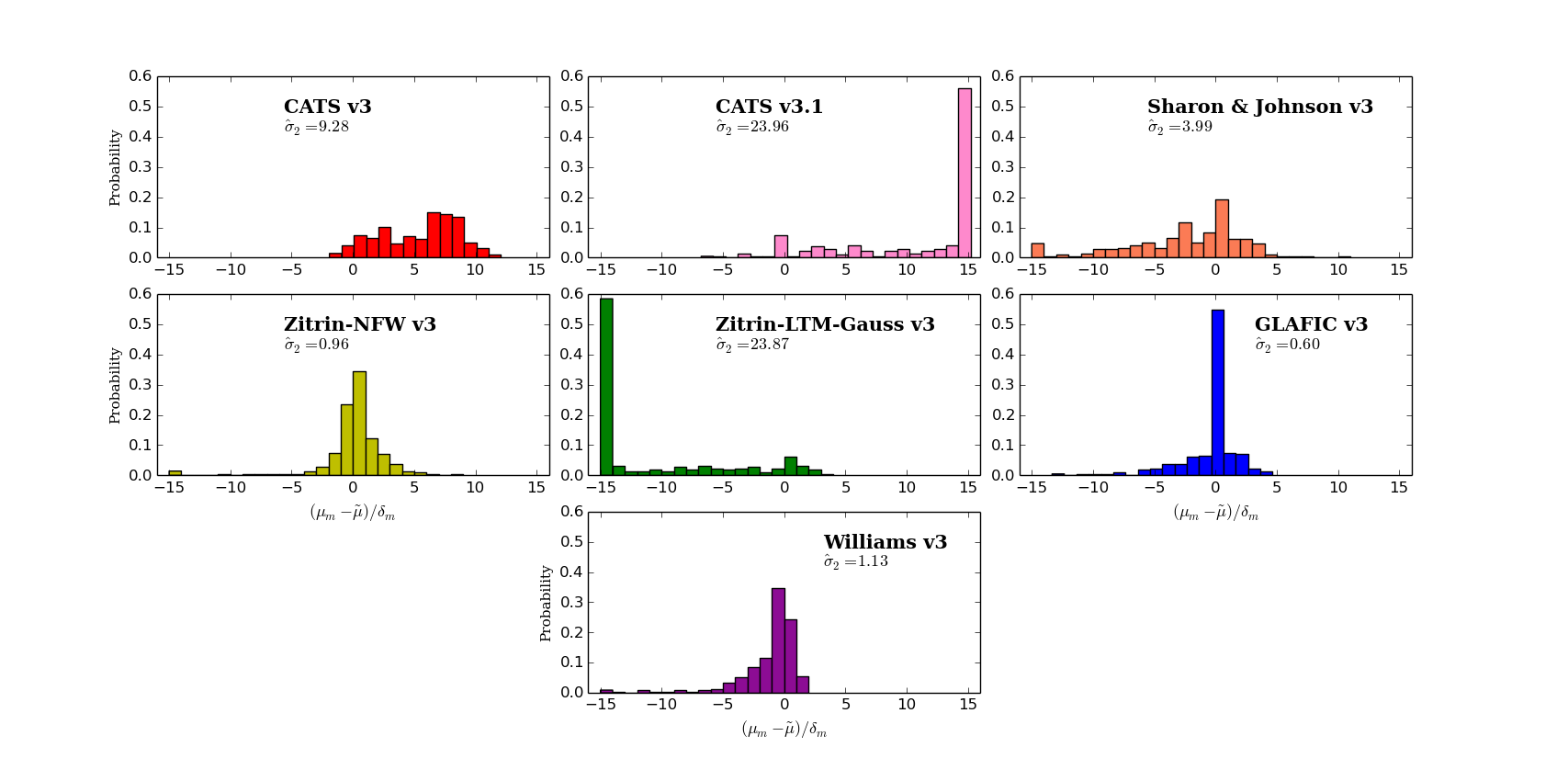}
\caption{Similar to Fig.~\ref{fig:Figure5} for Abell 2744, but using $(\mu_m-\tilde\mu)/\delta_m$ as the horizontal axis, i.e. the deviations from the median of all models is scaled by the quoted statistical uncertainties (68\% confidence limits) of each model at each sky location. Column 3 of Table~\ref{table2} shows the widths of these distributions (also displayed in each panel), quantified by the Normalized Median Absolute Deviation (see Section~\ref{comp}). Large (absolute) values indicate that the model's reported uncertainties are significantly less than the scatter among models.}\label{fig:Figure9}
\end{figure*}	

\begin{figure*}       %% figure 13
\includegraphics[width=1.0\linewidth, totalheight=0.4\textheight]{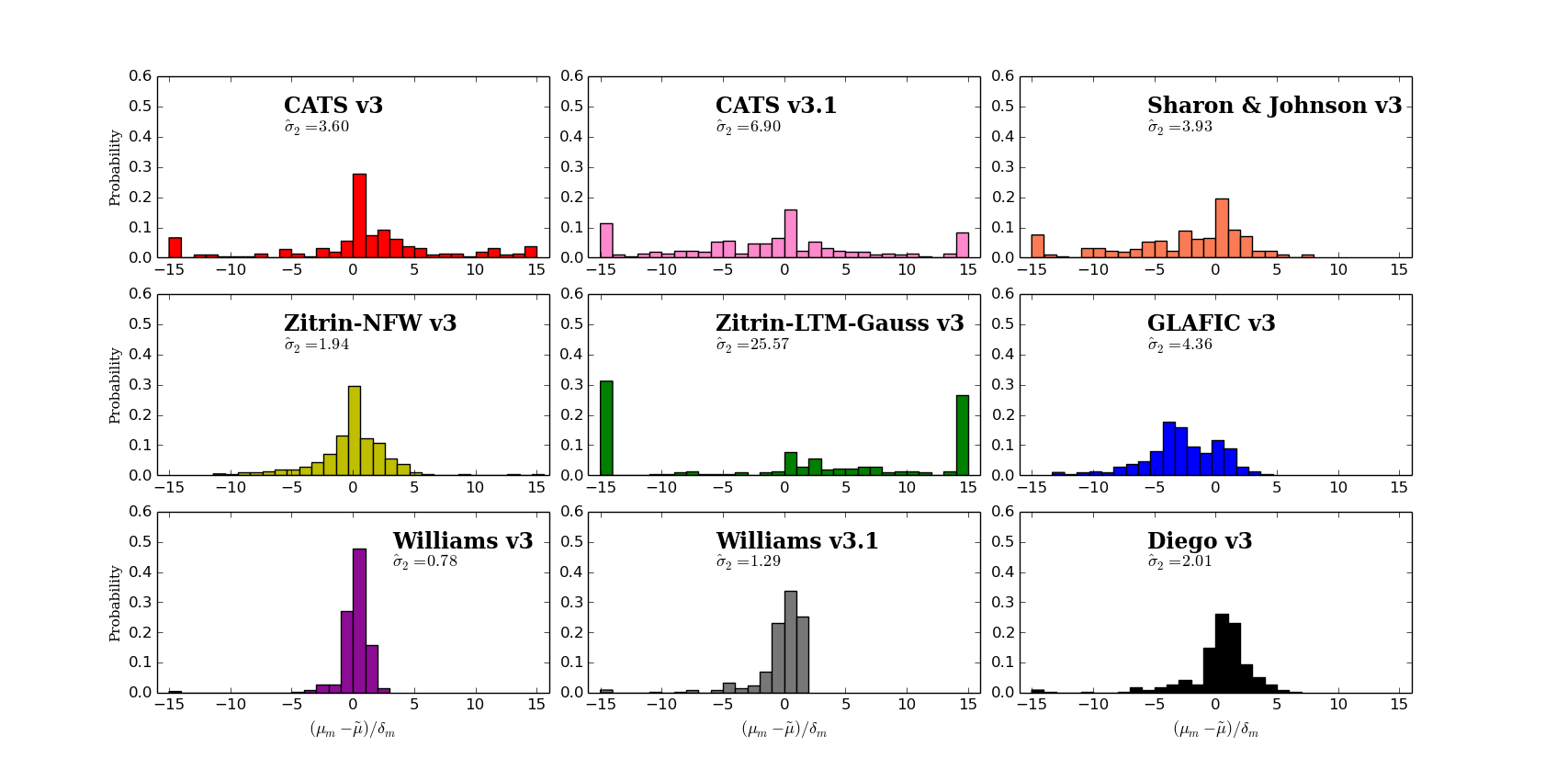}
\caption{Same as Fig.~\ref{fig:Figure9}, but for MACS J0416. Column 4 of Table~\ref{table2} shows the widths of these distributions (also displayed in each panel), quantified by the Normalized Median Absolute Deviation (see Section~\ref{comp}).}\label{fig:Figure10}
\end{figure*}

Each panel displays the rms width of the distribution, the Normalized Median Absolute Deviation, eq.~(\ref{eqNMAD}), where the median is over the sky locations that make up the histogram, and the fractional NMAD, fNMAD=NMAD/$\tilde\mu$. NMAD is less sensitive to outliers than rms, and is therefore a more robust measure of the width. These figures illustrate that when all models are considered, the bulk of their magnification predictions agree reasonably well, as judged by NMAD. However, the widths of the distributions get larger for higher magnifications: for $\tilde\mu$ around 2.5, 5, 10 and 15, NMAD values are, roughly, 0.7, 1.5, 4, and 8.5, respectively. 

The fNMAD values also show an increasing trend with $\tilde\mu$, and are plotted in Fig.~\ref{smfNMADvsmedian}, as function of the log of the median magnification, $\log{\tilde\mu}$, and using logarithmic bins of width $\pm$10\%. At $\tilde\mu\!<\!15$, MACS J0416 is better constrained than Abell 2744.

This overall representation makes one wonder how the individual models behave: is it just one model that gives rise to the widths of the distributions and the outliers or is there much variation between all models?  Each panel in Figs~\ref{fig:Figure5} and \ref{fig:Figure6} shows a histogram of $\mu_m - \tilde{\mu}$ for each model, for all sky locations. A narrow distribution around $\mu_m - \tilde{\mu}=0$ implies the model tracks the median of all models very well. A wide distribution implies large differences in this model compared to others. As in the previous two figures, models that have $|\mu_m - \tilde{\mu}|$ values beyond the box limits are piled up at the edges.

One of the conclusions that comes from Figs~\ref{fig:Figure5} and \ref{fig:Figure6} is that some models in some clusters tend to consistently over-predict magnifications compared to the median $\tilde\mu$ (e.g. CATS v3 and v3.1 in Abell 2744), while others under-predict (e.g. Sharon/Johnson and Zitrin-Gauss in Abell 2744). On the other hand, in MACS J0416 all models track $\tilde\mu$ well. We remind the reader that because our comparison benchmark is $\tilde\mu$, and not the true magnification, an over- or under-predicting model is not necessarily wrong!  ~This is also one of the messages of Fig.~\ref{fig:Figure1}, where most of the models over-predict the magnification at the location of HFF14Tom. However, in \cite{nor14} no tension between cluster mass models and three SN Ia magnification was found.

Figs~\ref{fig:Figure5} and \ref{fig:Figure6} ignore the models' magnification uncertainties. Figs~\ref{fig:Figure9} and \ref{fig:Figure10} are similar to the former, but now the horizontal axis shows $(\mu_m - \tilde{\mu})/\delta_m$, so the deviations from the median are scaled by each model's uncertainties. If the magnification fell below (above) $\tilde\mu$, the positive (negative) error was used for $\delta_m$. A narrow peaked histogram means a combination of two reasons: (i) the model's uncertainties are large, (ii) the model tracks the median of all models well. 

The widths of the distributions in Figs~\ref{fig:Figure5} and \ref{fig:Figure6}, $\hat\sigma_1$, for model $m$ are calculated as the Normalized Median Absolute Deviation between different sky locations $\vec\theta$, given by eq.(\ref{eqNMAD}), where the median is taken over $N=214$ sky locations. These values are presented in columns 2 and 4 of Table~\ref{table2} for Abell 2744 and MACS J0416, respectively. We chose to present these instead of standard deviations because NMAD are less sensitive to outliers. The widths of the distributions in Figs~\ref{fig:Figure9} and \ref{fig:Figure10} are calculated using an equation similar to the above, $\hat\sigma_2={\rm NMAD}/\delta_m(\vec\theta)$, and are shown in columns 3 and 5 of the same table. The table also shows that even if, on average, a given model follows the median of all models well (as most models do in Fig.~\ref{fig:Figure10}), magnifications at some sky locations are still very discrepant.

Figs~\ref{fig:Figure9} and \ref{fig:Figure10} reveal that the histograms for Abell 2744 (Fig.~\ref{fig:Figure9}) are noticeably less symmetric than those for MACS J0416 (Fig.~\ref{fig:Figure10}), implying that the models for Abell 2744 tend to disagree with each other more than those for MACS J0416. Because the lens inversion methods are the same for the two clusters (at least the 7 models in common) the difference must arise from something else, for example (i) the number of images is smaller in Abell 2744 ($\sim 60$ vs. $\sim 90$), (ii) the quality of lensing constraints is weaker (e.g., a lower fraction have spectroscopic redshifts), (iii) the morphology of Abell 2744 is more prone to lensing degeneracies. We return to this in Section~\ref{degen}.

\begin{figure*}    %% figure 14
\includegraphics[width=1.1\linewidth, totalheight=0.38\textheight]{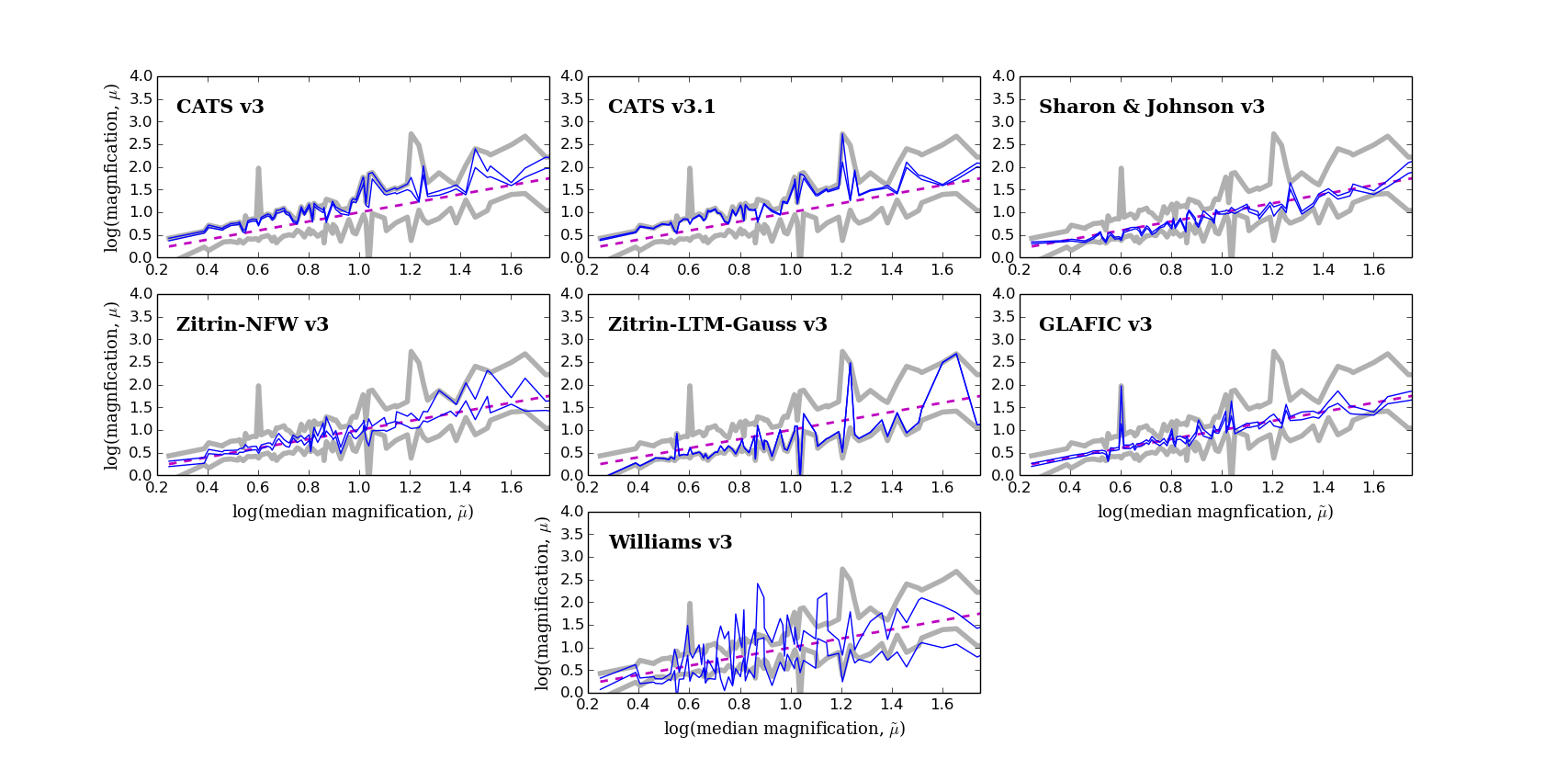}
\caption{Comparison of magnification uncertainties from each model against the full uncertainty range from all models, for a subset of sky locations. Each panel shows the comparison for a single model, as labeled. In all panels, each column ($x$ coordinate) corresponds to 1 out of 65 sky locations, drawn from the 214 points marked with blue points in Figure~\ref{ds9}. For convenience, these points are sorted in order of increasing median magnification value, $\tilde\mu$.  Along any given column, the vertical space between the two blue lines corresponds to the error bar (68\% confidence range) of the given model, for a single point on the sky. The vertical space between the gray lines at each $x$ position represents the full confidence range from the union of all models (excluding the Williams/\grale~ model).  That is, the gray lines (identical in all panels) trace out the maximum (minimum) value from the set of upper (lower) confidence range values of all models (such as shown in Figures~\ref{fig:Figure3} and \ref{fig:Figure4}), except the Williams/\grale~ model. }\label{fig:Figure13}
\end{figure*}

\begin{figure*}       %% figure 15
\includegraphics[width=1.1\linewidth, totalheight=0.38\textheight]{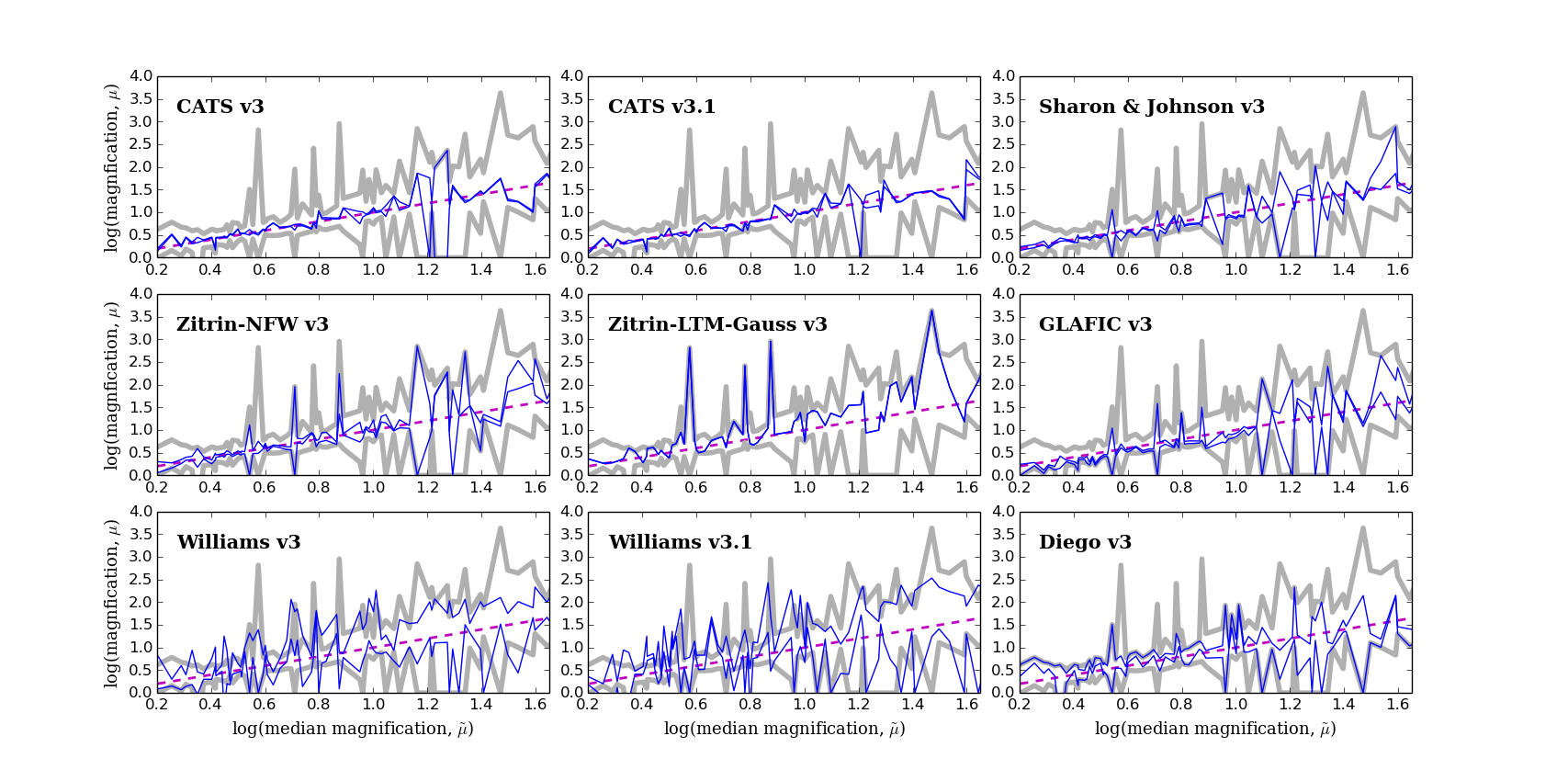}
\caption{Same as Figure~\ref{fig:Figure13}, but for MACS J0416.}\label{fig:Figure14}
\end{figure*}

Finally, in Figs~\ref{fig:Figure13} and \ref{fig:Figure14} we take a closer look at the range of magnification uncertainties of the two clusters, with the aim of estimating the systematic uncertainties in the reconstruction, which we assume to be reasonably represented by the full range of statistical uncertainties quoted for each model. As stated in the introduction, this assumption is justified because the source/image inputs are of the same quality and similar quantity in all models, and the resulting lens plane rms are also similar in all models (see Table~\ref{SLdata}). The horizontal axis in all the panels of these two figures shows the median magnification of all models, $\tilde\mu$. We place these in order of increasing $\tilde\mu$, for clarity. The two thin blue lines in each panel represent a given model's 68\% confidence range. The two thick gray lines (same in all panels) show the maximum and minimum of all models' 68\% range, except for Williams/\grale. The latter model was singled out because it has the largest uncertainties at most sky locations, as was already illustrated in Figs~\ref{fig:Figure1}, \ref{fig:Figure3} and \ref{fig:Figure4}. Including Williams/\grale~ results would incorrectly suggest that their uncertainties match the systematic errors. 

The bands spanned by the thick gray lines can be interpreted as the systematic uncertainties. As was already indicated by earlier figures, some models tend to be systematically below $\tilde\mu$, while some are above. It is clear that most models' error-bars are smaller than the extent of all the models' predictions taken as a whole.  There are sets of models whose predictions lie completely outside of each other's error-bars, like CATS and Zitrin-LTM-Gauss in Abell 2744. The model which comes closest to estimating systematic errors is Williams/\grale~ v3, for both Abell 2744 and MACS J04616. However, it does not do so perfectly, as Figs~\ref{fig:Figure3} and \ref{fig:Figure4} illustrate.

The information in Figs~\ref{fig:Figure13} and \ref{fig:Figure14} can be reframed quantitatively as follows. For any two models, 1 and 2, we calculate how far their predictions are from each other, in terms of their 68\% confidence range:
 ${\rm median}\Bigl[(\mu_1-\mu_2)/[0.5(\delta_1+\delta_2)]\Bigr]$, where the median is taken over $N=214$ sky locations, and the ordering of 1 and 2, and the use of upper vs. lower error-bars depends on which of the two magnifications is larger at a given sky location. These are tabulated in Table~\ref{table3}: the two values on either side of the vertical bar (in each cell in the table) are for Abell 2744 and MACS J0416, respectively. In most cases, the smaller values mean that the quoted uncertainties in either model 1 or 2 are large, making the difference between models small.

\begin{table}
\centering
\begin{tabular}{|c|c|c|c|c|}
        \hline  & \multicolumn{2}{c}{Abell 2744} & \multicolumn{2}{c}{MACS J0416}\\
	\hline Models           & $\hat\sigma_1$ & $\hat\sigma_2$ & $\hat\sigma_1$ & $\hat\sigma_2$ \\ 
  & (Fig.~\ref{fig:Figure5}) & (Fig.~\ref{fig:Figure9}) & (Fig.~\ref{fig:Figure6}) & (Fig.~\ref{fig:Figure10}) \\
        \hline
	\hline CATS v3          & 7.81   & 9.28  & 0.67  & 3.60 \\ 
	\hline CATS v3.1        & 7.19   & 23.96 & 0.74  & 6.90 \\ 
	\hline Sharon v3        & 1.50   & 3.99  & 1.08  & 3.93 \\ 
	\hline Zitrin-NFW v3    & 1.22   & 0.96  & 1.10  & 1.94 \\ 
	\hline Zitrin-Gauss     & 4.28   & 23.87 & 2.95  & 25.57\\ 
	\hline GLAFIC v3        & 0.39   & 0.60  & 1.76  & 4.36 \\ 
	\hline Williams v3      & 3.74   & 1.13  & 3.10  & 0.78 \\ 
	\hline Williams v3.1    & --     & --    & 3.93  & 1.29 \\ 
	\hline Diego v3         & --     & --    & 2.96  & 2.01 \\ 
	\hline 
\end{tabular} 
\caption{The Normalized Median Absolute Deviation (NMAD), $\hat\sigma_1$, and NMAD divided by reported uncertainties, $\hat\sigma_2$ (see Section~\ref{comp}) of the histograms plotted in Figs~\ref{fig:Figure5} and \ref{fig:Figure6} (columns 2 and 4) and Figs~\ref{fig:Figure9} and \ref{fig:Figure10} (columns 3 and 5), for Abell 2744 and MACS J0416. The large values indicate that some individual magnifications are very different from $\tilde\mu$, and would likely lie outside of the boundaries of the corresponding figures. The reason why some models have larger $\hat\sigma_i$ values for Abell 2744 than for MACS J0416 is likely to be the approximate mass sheet degeneracy (see Section~\ref{degen}).}\label{table2}
\end{table}

\begin{figure}    %% figure 16
\centering
\includegraphics[width=0.95\linewidth]{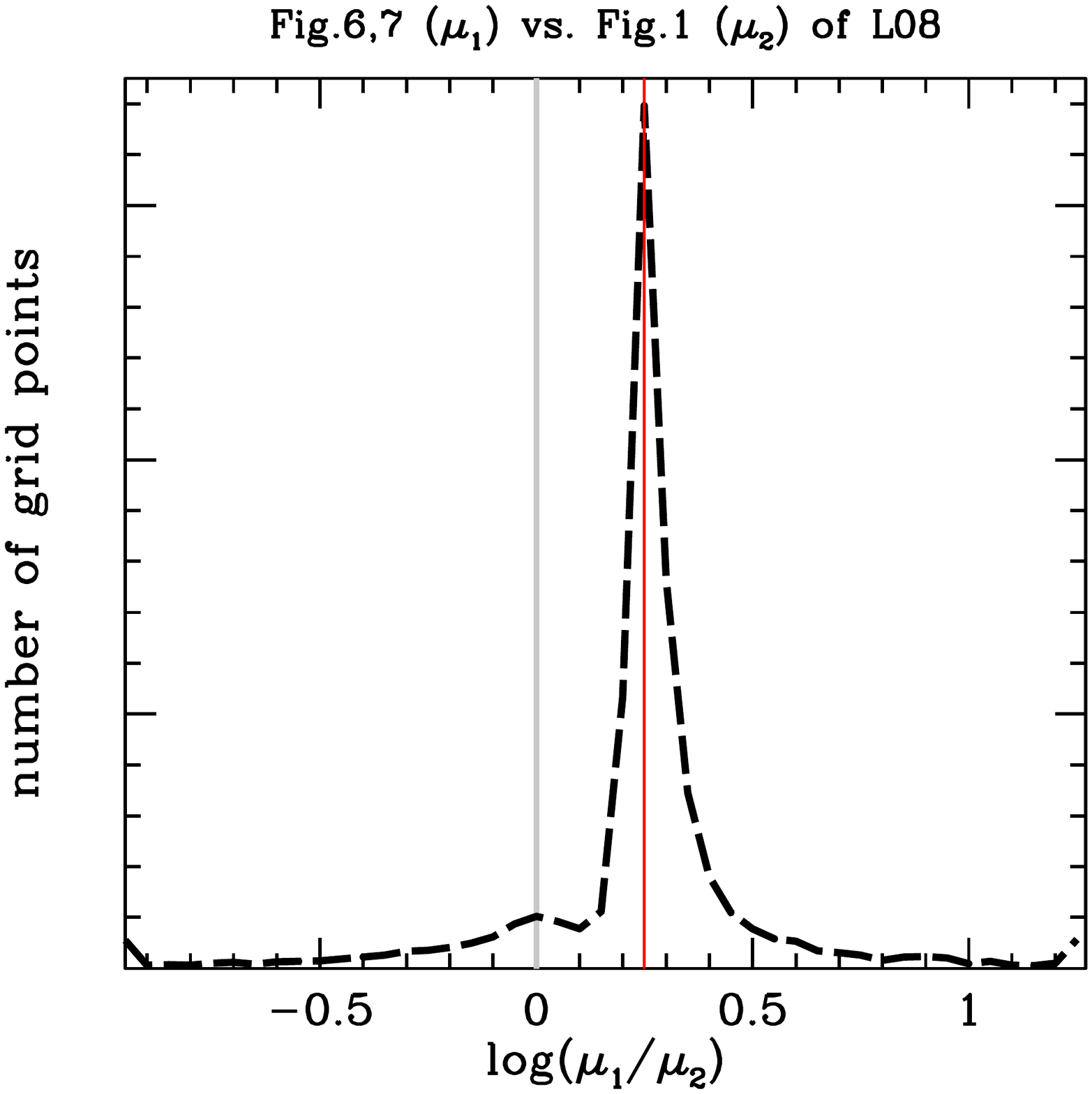}
\vskip-0.7in
\caption[]{A demonstration of an approximate Mass Sheet Degeneracy. The two models, 1 and 2, presented in \cite{lie08} are MSD-degenerate. The black dashed histogram shows the distribution of $\log(\mu_1/\mu_2)$, taken over many locations,  placed on a regular grid within the mass distribution. The scaling $s$ used to generate model 1 from model 2 is $s=0.75$, but the projected mass density of the added sheet varies with position, such that the two mass models reproduce the image positions exactly, even though the two sources are at different redshifts. The red vertical line is at $-\log(s^2)=0.2499$. If the histogram were centred on $\log(\mu_1/\mu_2)=0$ (gray vertical line) that would indicate no MSD. See Section~\ref{degen}.}
\label{smjori}
\end{figure}

\section{Why models' magnifications differ: lensing degeneracies}\label{degen}

The previous sections have shown that magnification predictions from any given sky location differ between models, often by more than the stated uncertainties. Since all models reproduce observed images positions well---lens plane rms is typically $<\!1^{\prime\prime}$---the differences between magnification maps are largely due to lensing degeneracies, i.e. a given distribution of observed images can be reproduced by many different mass distributions. 

The most recognized degeneracy is mass sheet degeneracy, MSD \citep{gsf88,saha00}. It rescales the mass surface density by $s$: $\kappa(\vec\theta)\rightarrow s\kappa(\vec\theta)$, and adds a constant thickness mass sheet of surface mass density $(1-s)$, in units of critical density for lensing. The source locations are transformed as $\vec\beta\rightarrow s\vec\beta$, and hence all magnifications are scaled by $s^2$. This global exact degeneracy, which we will call classic MSD, is broken in all HFF clusters because of sources at multiple redshifts. Another exact global degeneracy is the source-plane transformation, or SPT \citep{ss13,ss14,unr16}, which generalizes MSD by replacing constant $s$ with a function of source location, $\vec\beta$. As a result, the image magnification ratios are not affected, but actual magnifications are, by different amounts. Because it affects axisymmetric lenses only, the exact SPT is broken in all HFF clusters. 

However, in this section we demonstrate that other degeneracies \citep{lie12} are not broken. This includes the generalized MSD, where the scaling, $s$, of the mass and the added mass sheet vary somewhat as a function of location within the cluster. Hereafter we use MSD to mean generalized MSD. Note that the generalized MSD still allows for multiple lens model solutions that reproduce the positions of images exactly. If that condition is loosened somewhat, allowing for small positional differences between observed and model generated images, the range of degenerate lens solutions becomes even wider. Both exact and approximate generalized MSD operate in real cluster reconstructions. Approximate SPT \citep{ss14} may also be present in mass reconstructions, and affect the magnifications of images. Note that MSD and SPT, and especially their approximate versions may be the same transformation in some cases. We will not attempt to differentiate between them, and will concentrate on how these---and possibly other---degeneracies affect magnification predictions of the various lens models. A recent paper examining another HFF cluster, MACS J0717.5+3745 \citep{lim16} concluded that because of degeneracies one cannot differentiate between cuspy and cored \lenstool-based parametric models of the cluster, and that the differences in magnifications obtained by various reconstruction methods using pre-HFF data can be considerable.

The diagnostic we will use is $\log(\mu_1/\mu_2)$, where $\mu_1$ and $\mu_2$ are magnifications at the same sky locations of two different models, so the comparison is between any two models. This quantity can also be thought of as quantifying how much two models deviate in their predicted magnifications.

Because magnifications can become very large near critical lines (formally infinite for point sources), some care has to be taken when using our proposed metric. We carried out our analysis after eliminating all sky locations where {\em any one} of the models exceeded some $\mu_{max}$. Results with varying $\mu_{max}$ between 50 and 500 are very similar (within Poisson noise), so below we quote results for $\mu_{max}=200$, which reduces the original number of sky locations by 5\%-10\%. (To facilitate comparison between Abell 2744 and MACS J0416, we renormalized the results below as if the same number of sky locations were used.)

\begin{figure*}    %% figure 17
\centering
\includegraphics[width=.32\textwidth]{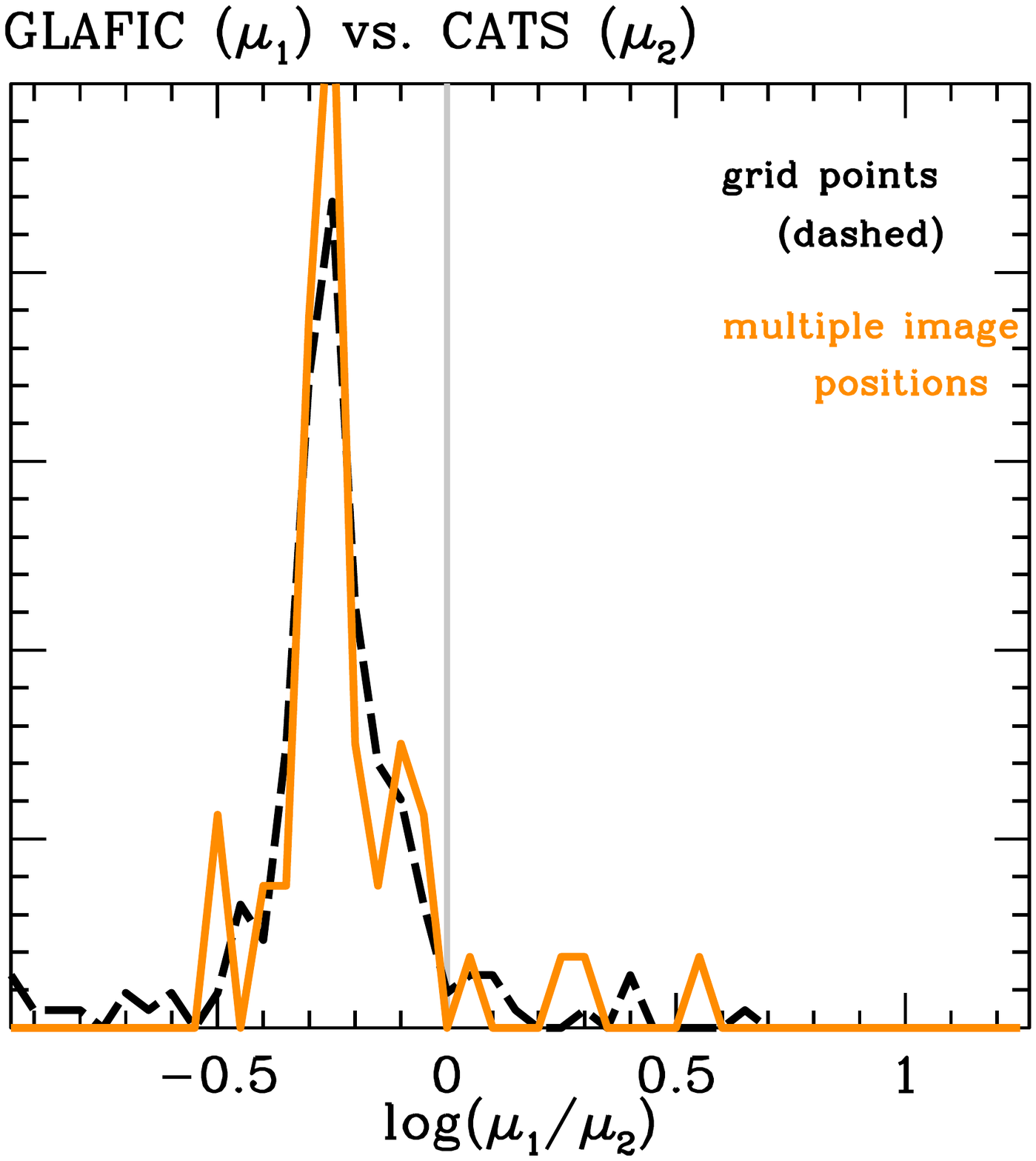}
\vspace{-27pt}
\includegraphics[width=.32\textwidth]{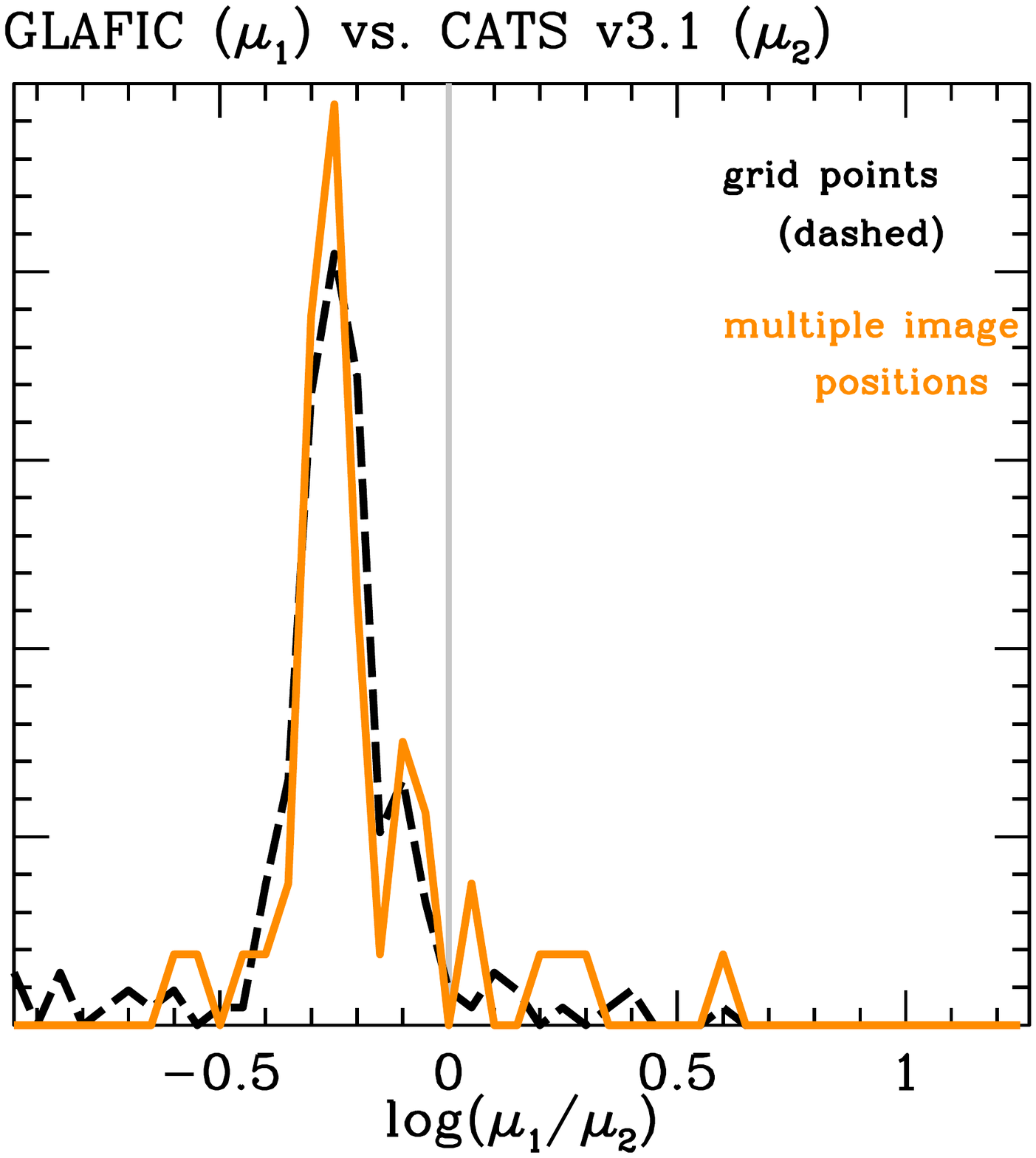}
\vspace{-27pt}
\includegraphics[width=.32\textwidth]{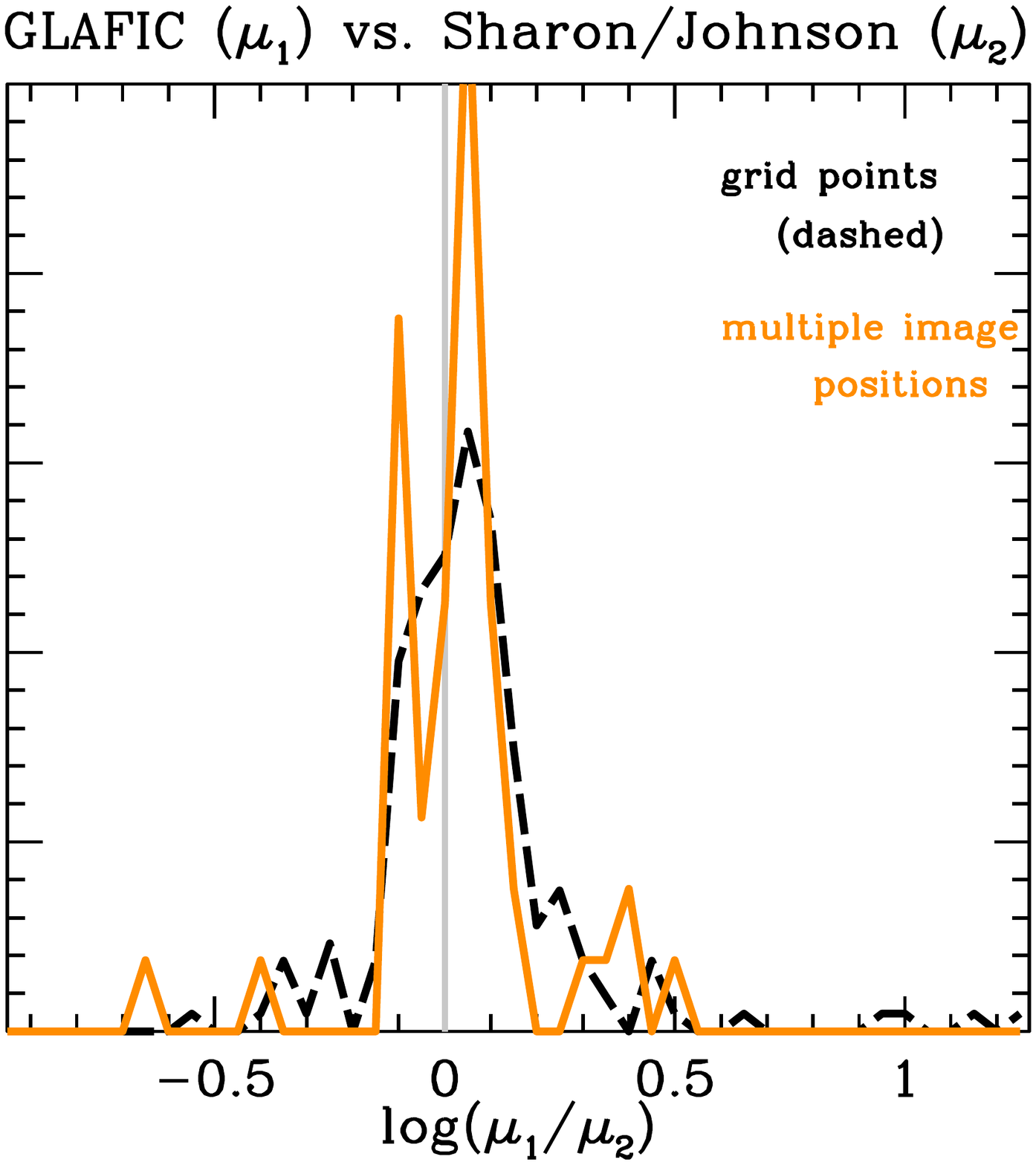}
%\hfill
\includegraphics[width=.32\textwidth]{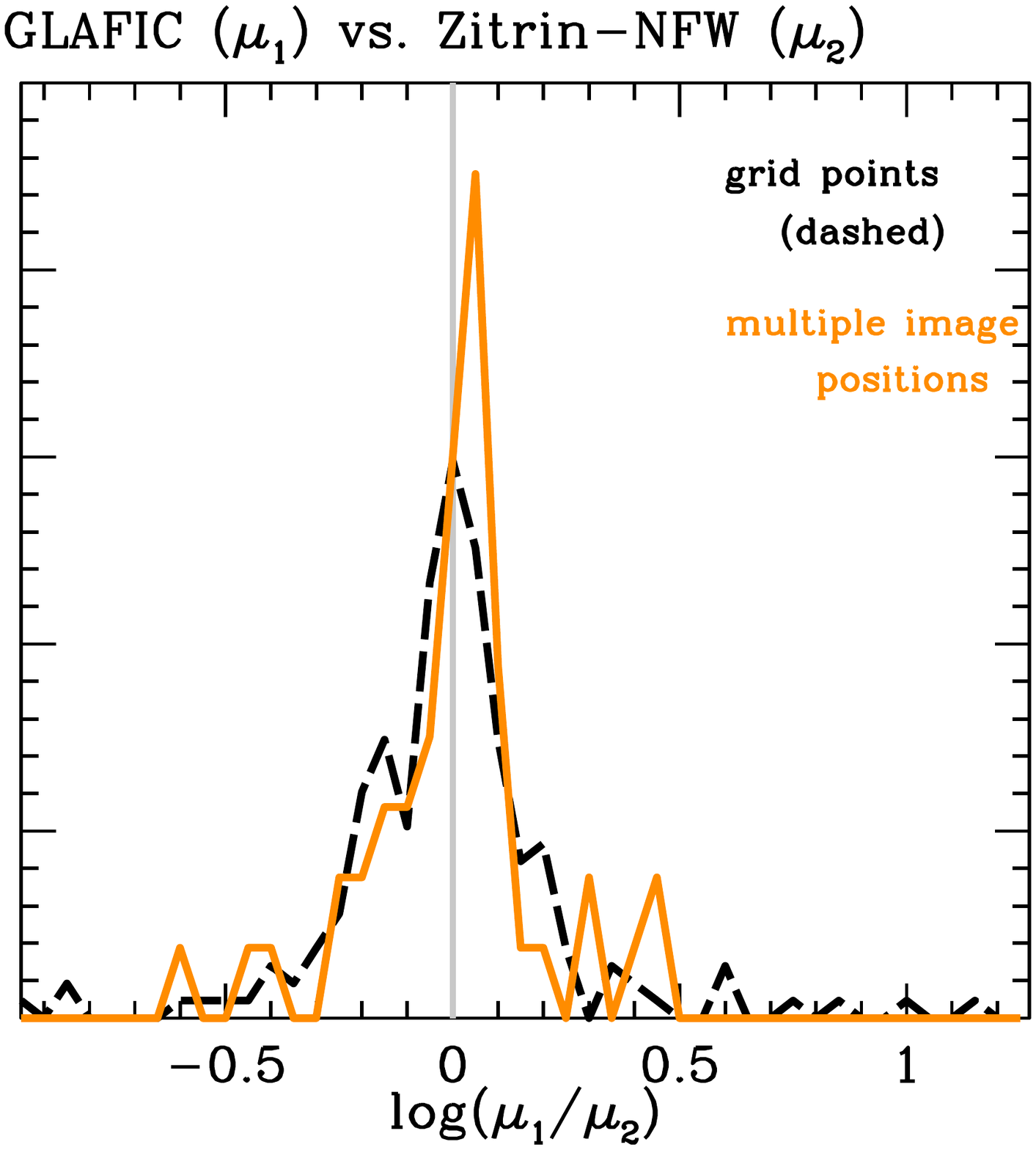}
\vspace{-27pt}
\includegraphics[width=.32\textwidth]{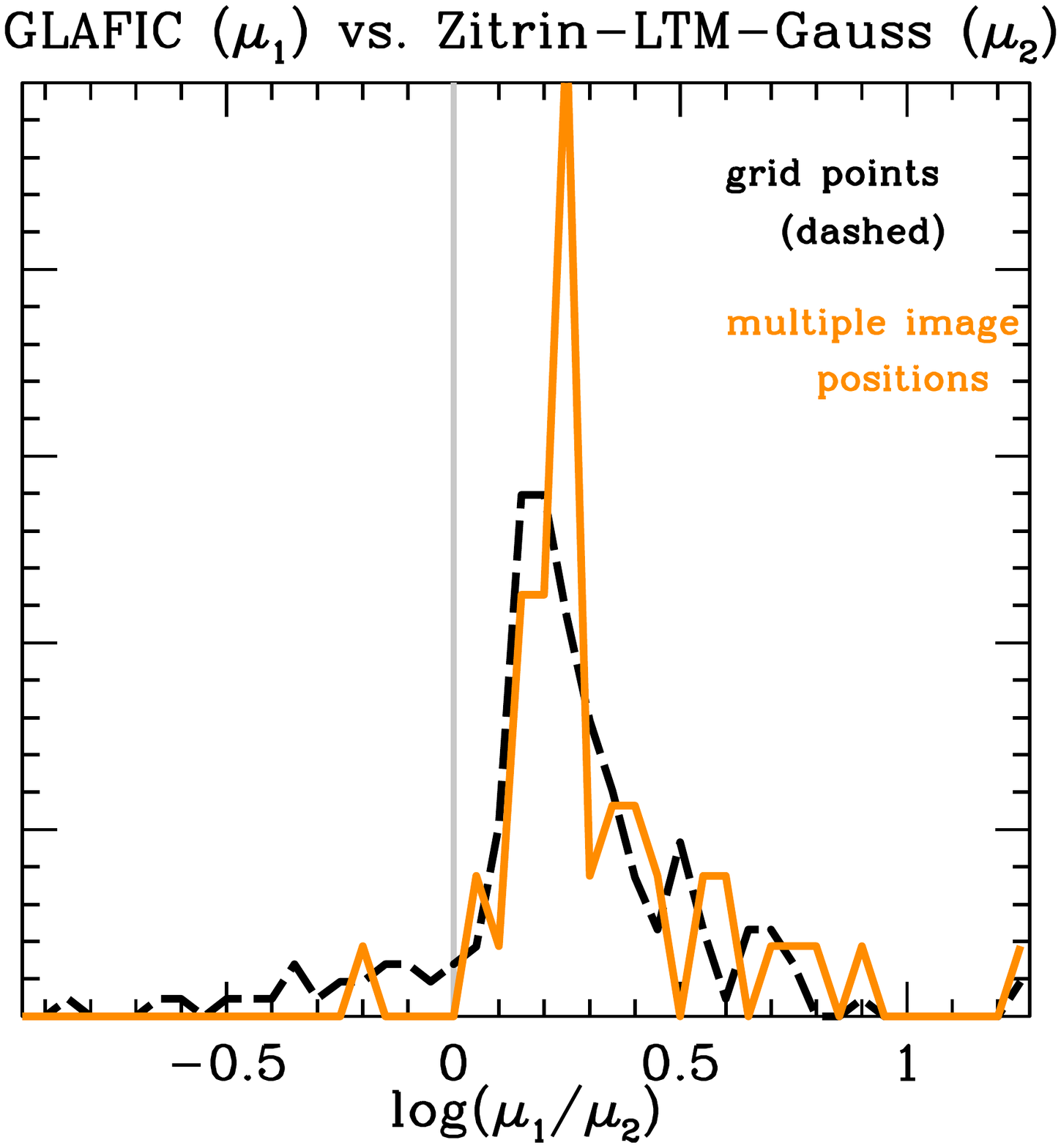}
\vspace{-27pt}
\includegraphics[width=.32\textwidth]{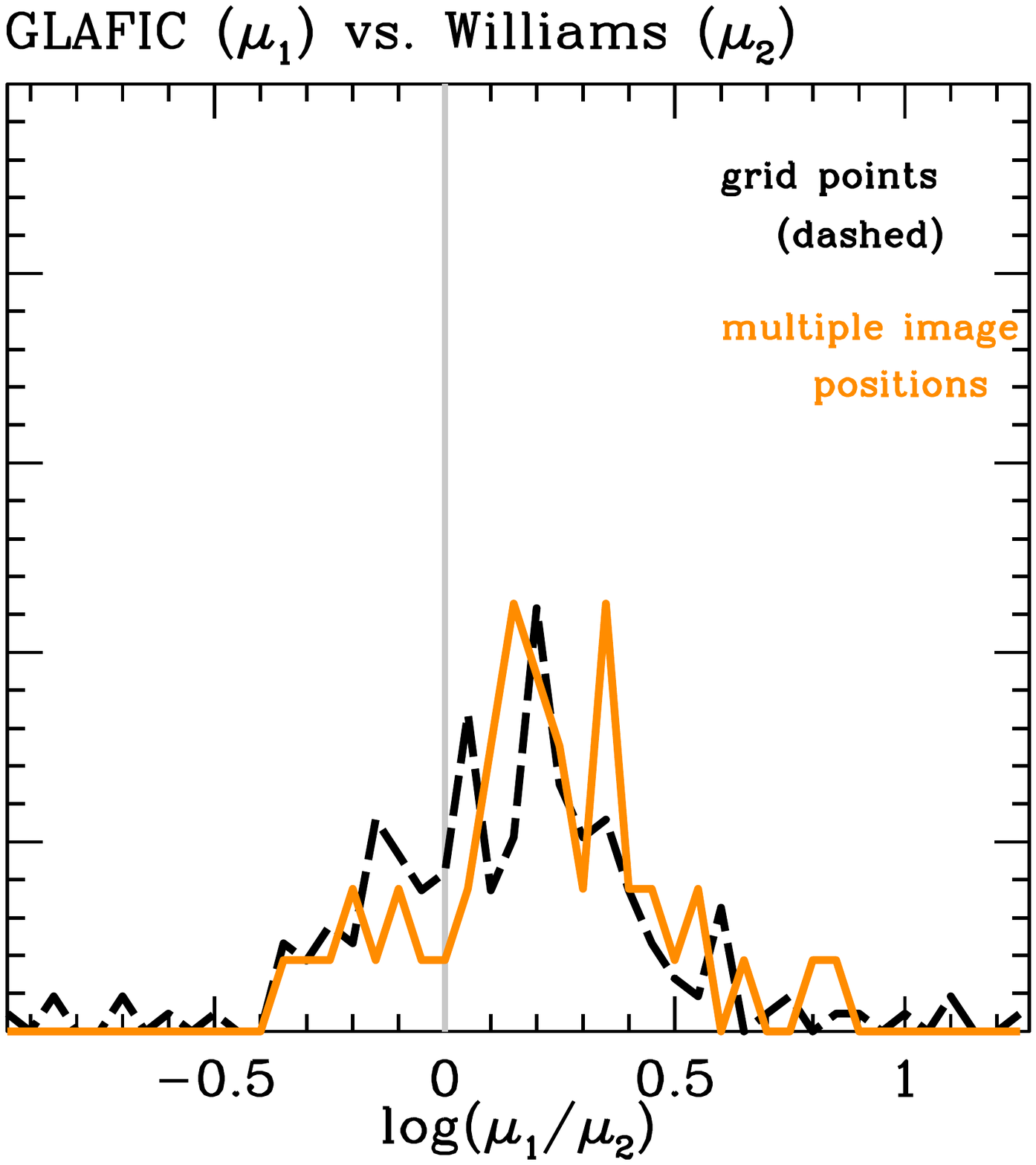}
\vspace{+25pt}
\caption{Magnification correlations for six pairs of models of Abell 2744. Each panel compares two models. Model 1 is always GLAFIC v3, while model 2 is as indicated above each panel. All models are v3 unless indicated otherwise. (These plots are similar to Fig.~\ref{smjori}.) The black dashed lines are histograms of $\log(\mu_1/\mu_2)$ distributions, and magnifications are taken at the grid sky locations shown as blue points in Fig.~\ref{ds9}. The orange solid line histogram is similar, but uses magnifications at the observed GOLD+SILVER multiple images (red squares in Fig.~\ref{ds9}).}
\label{fig:magcorr2744}
\end{figure*}

\begin{figure*}    %% figure 18
\centering
\includegraphics[width=.32\textwidth]{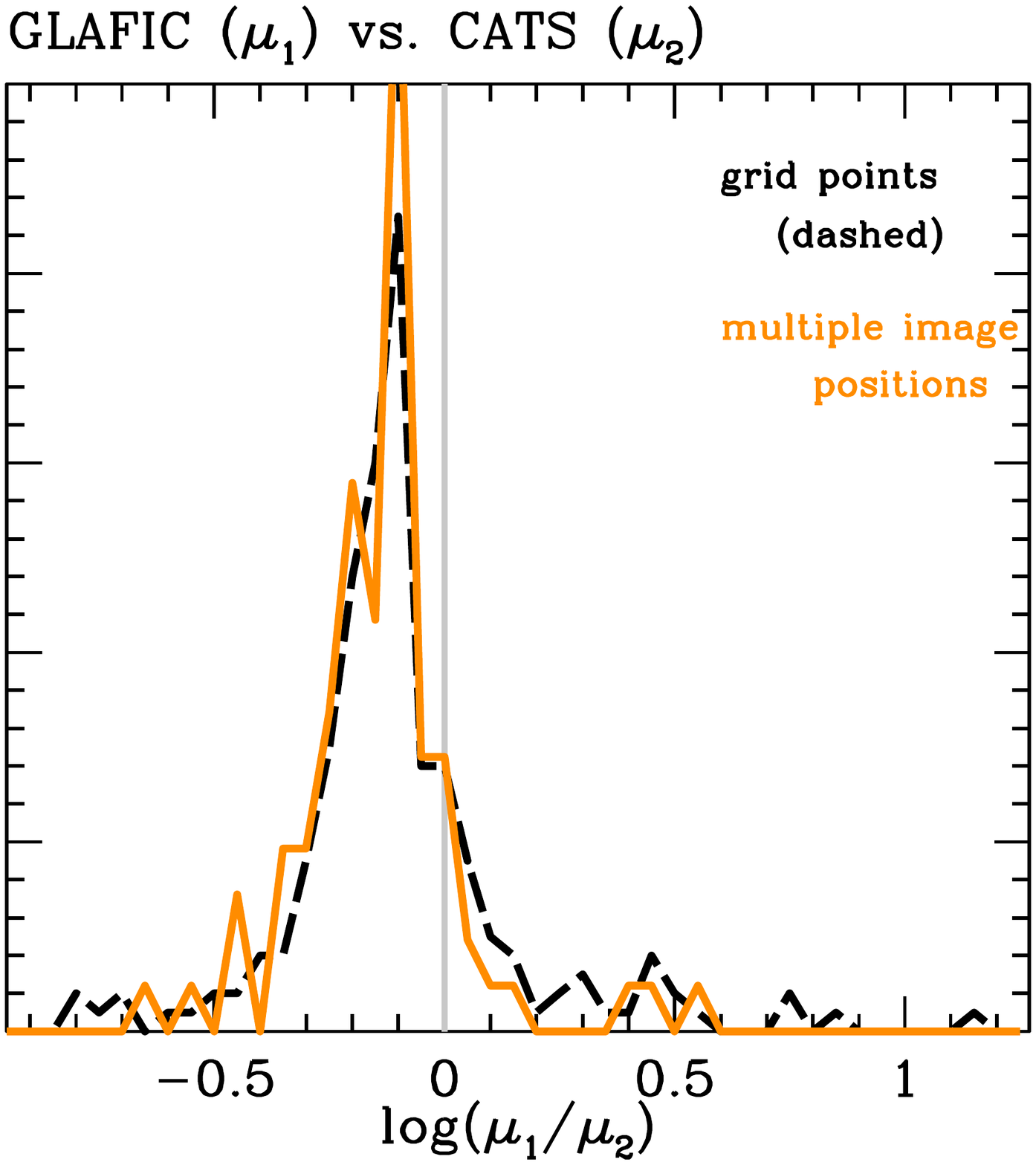}
\vspace{-27pt}
\includegraphics[width=.32\textwidth]{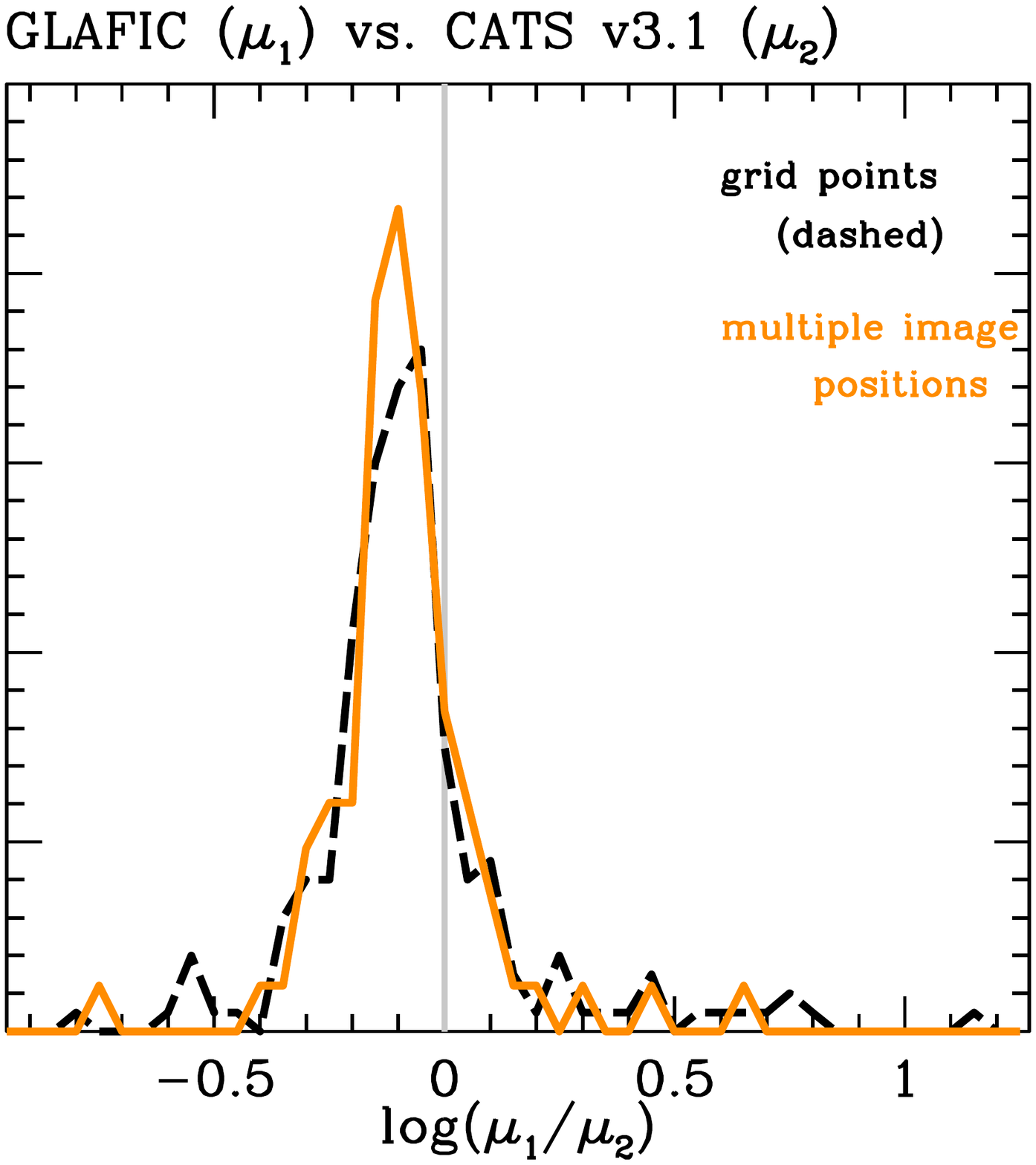}
\vspace{-27pt}
\includegraphics[width=.32\textwidth]{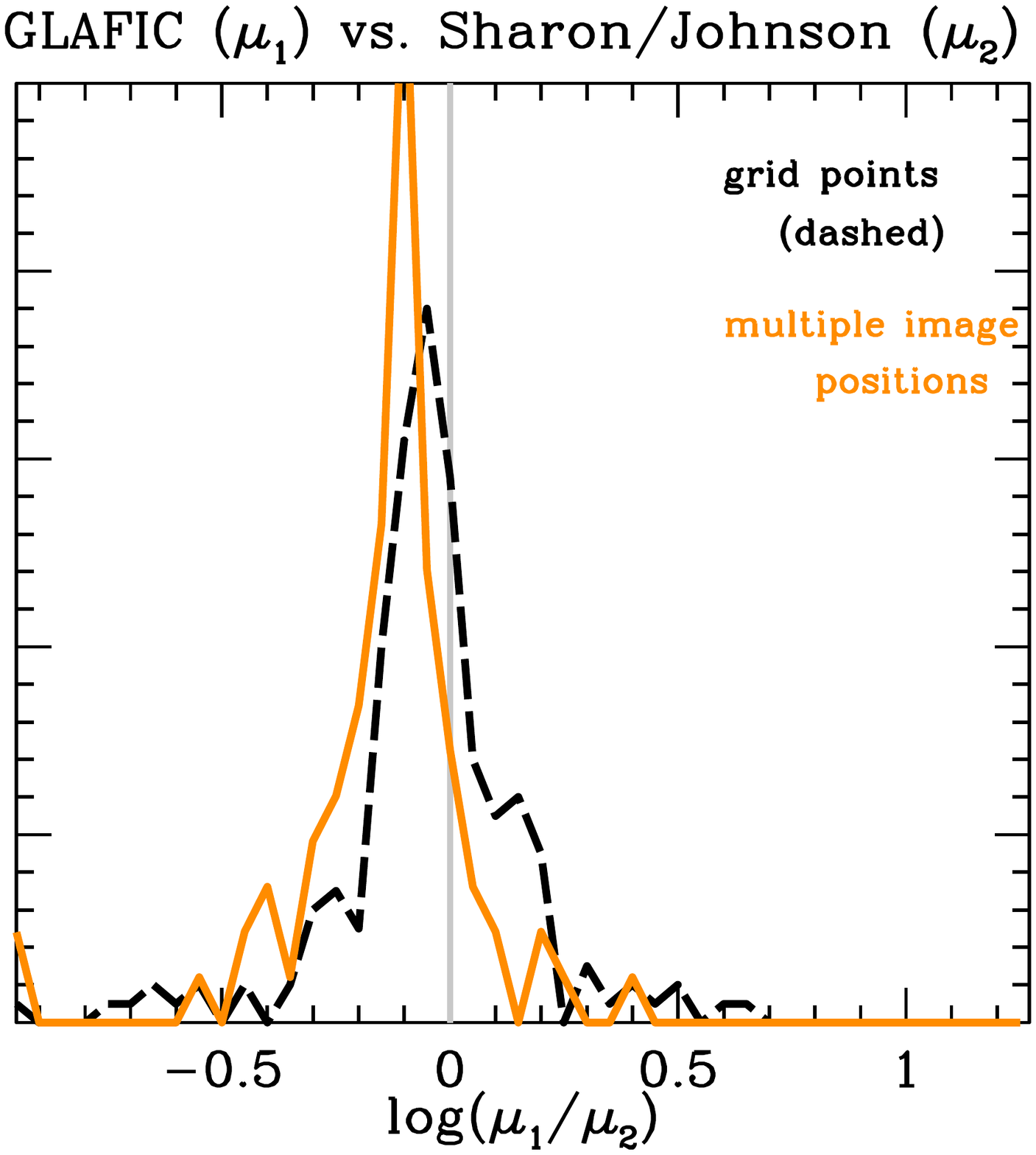}
%\hfill
\includegraphics[width=.32\textwidth]{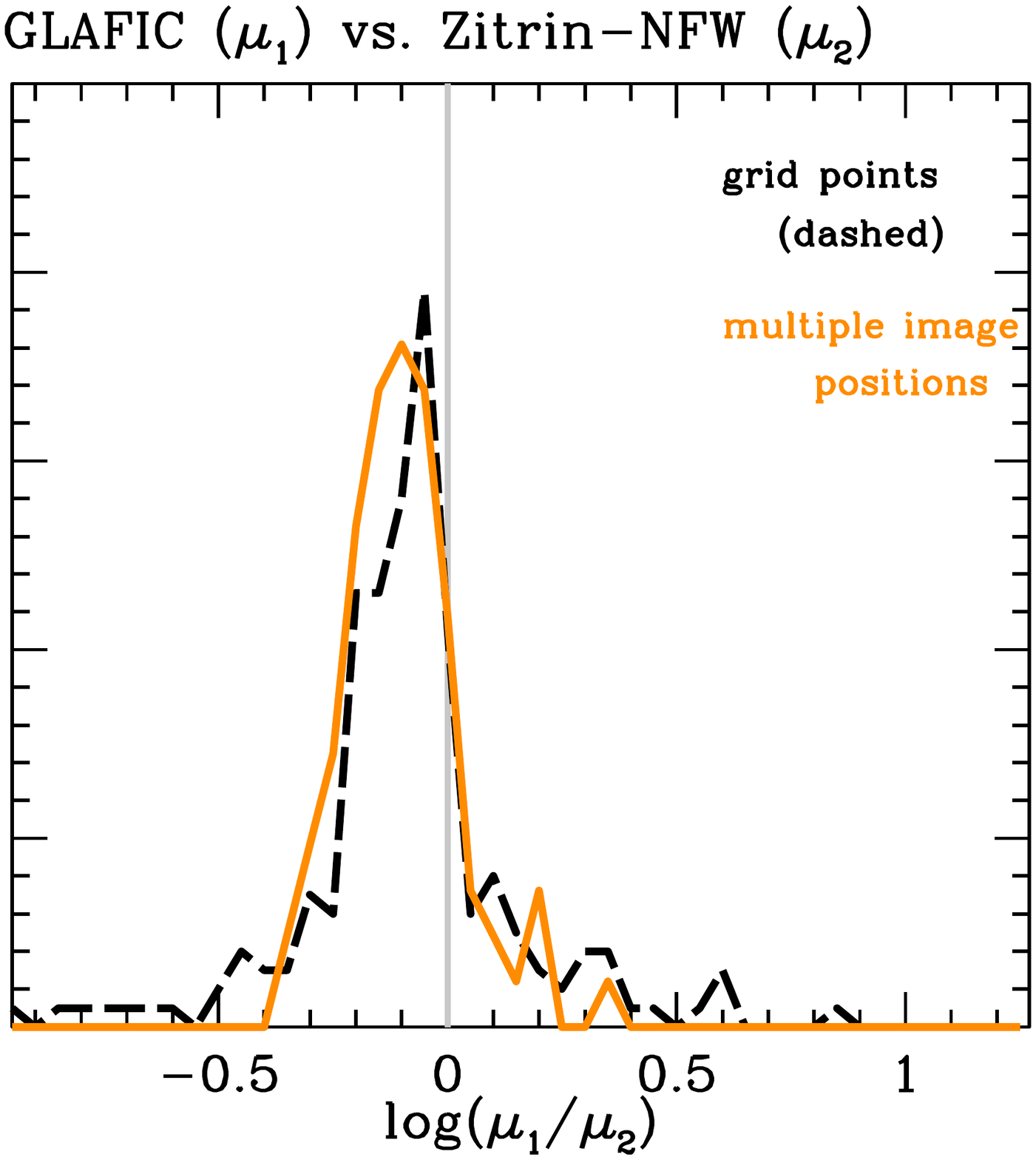}
\vspace{-27pt}
\includegraphics[width=.32\textwidth]{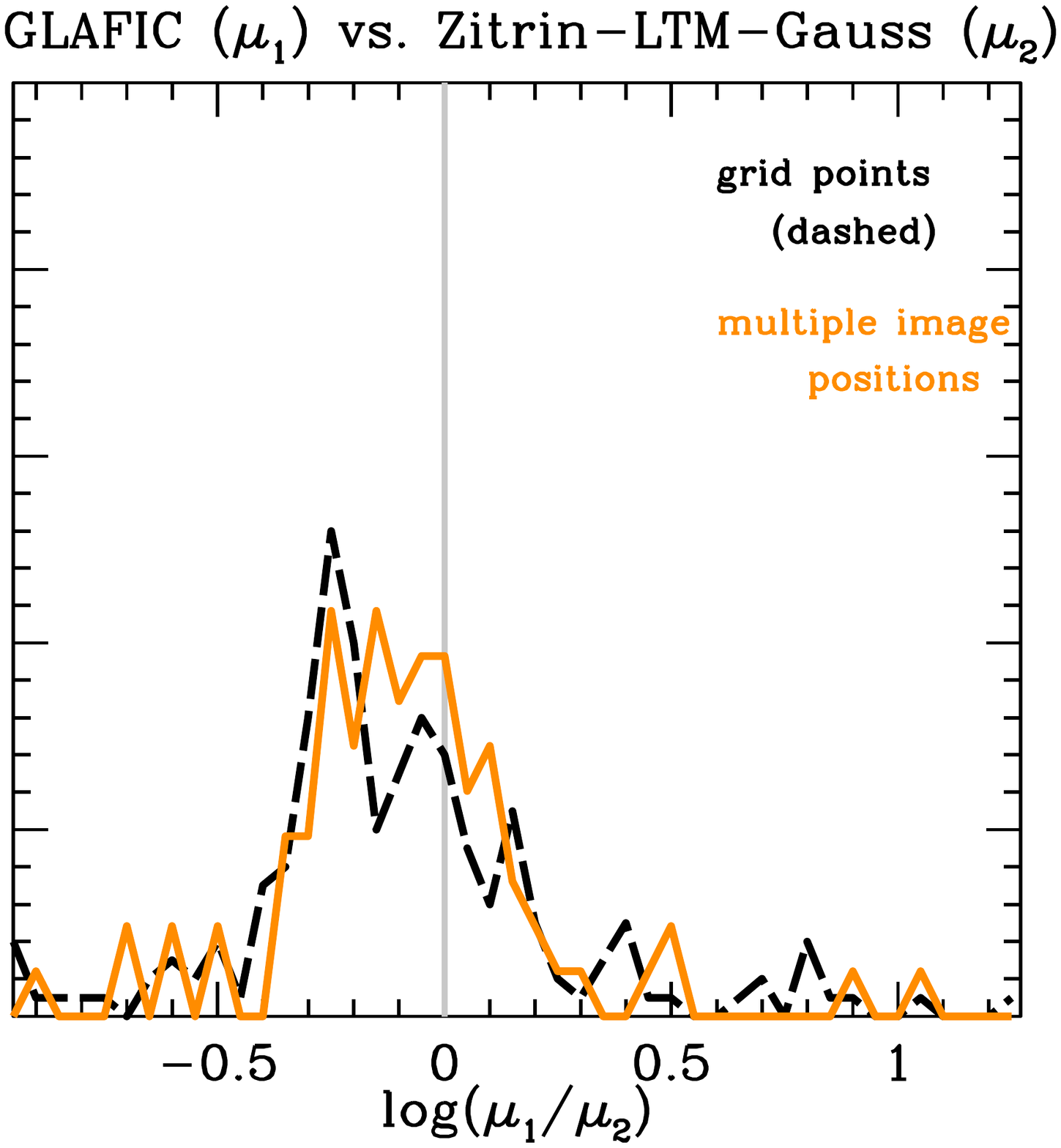}
\vspace{-27pt}
\includegraphics[width=.32\textwidth]{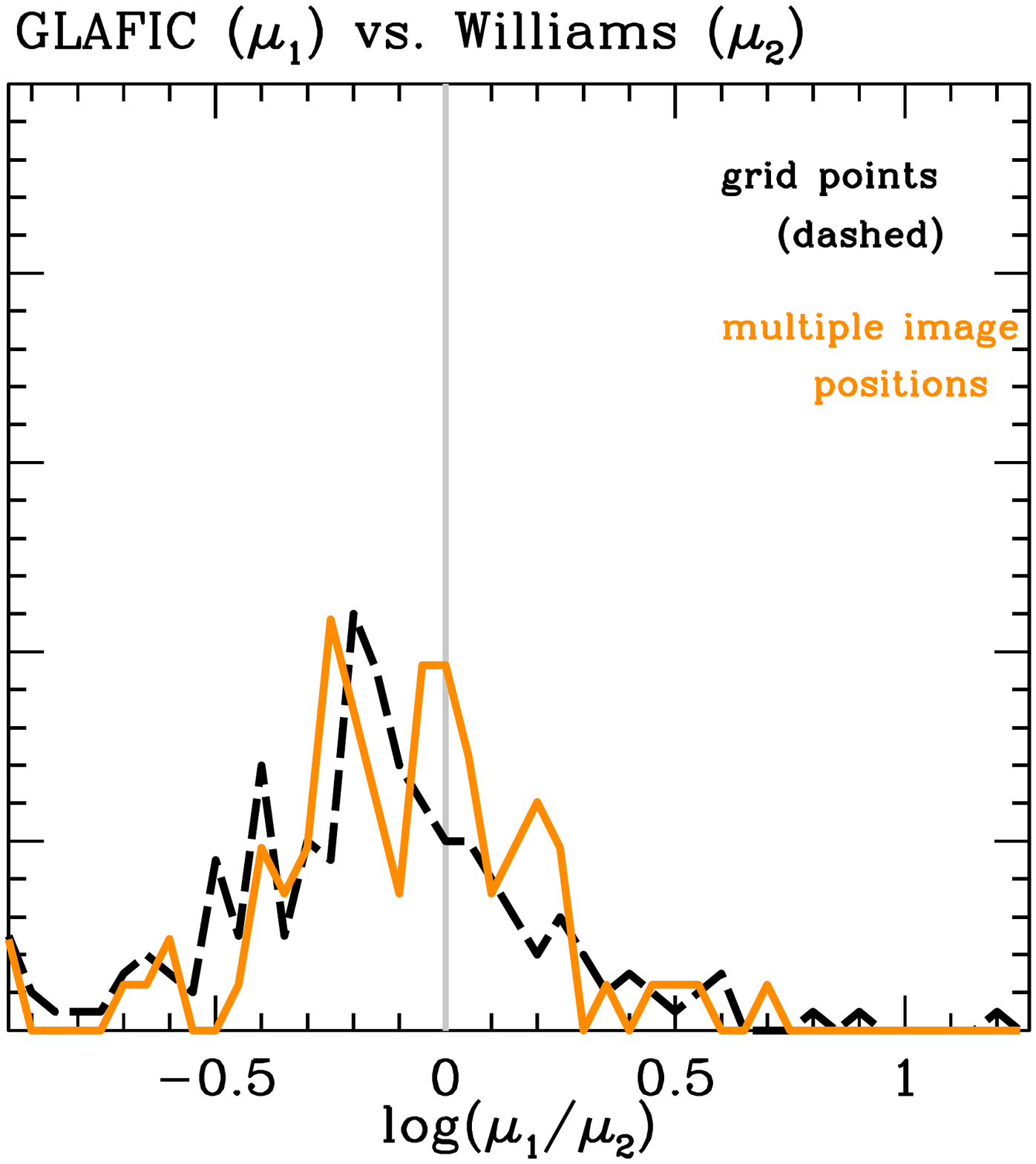}
%\hfill
\includegraphics[width=.32\textwidth]{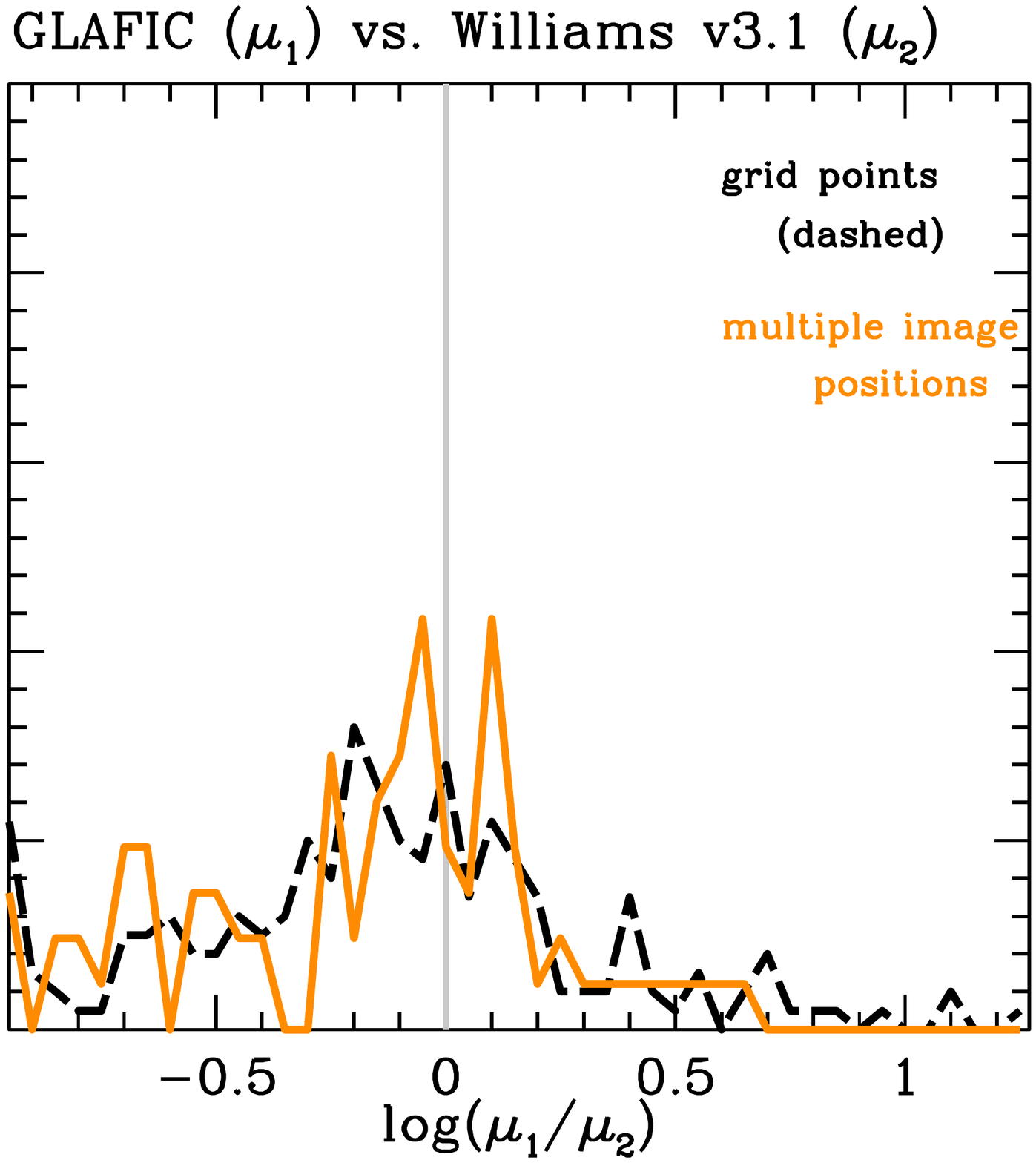}
\vspace{-27pt}
\includegraphics[width=.32\textwidth]{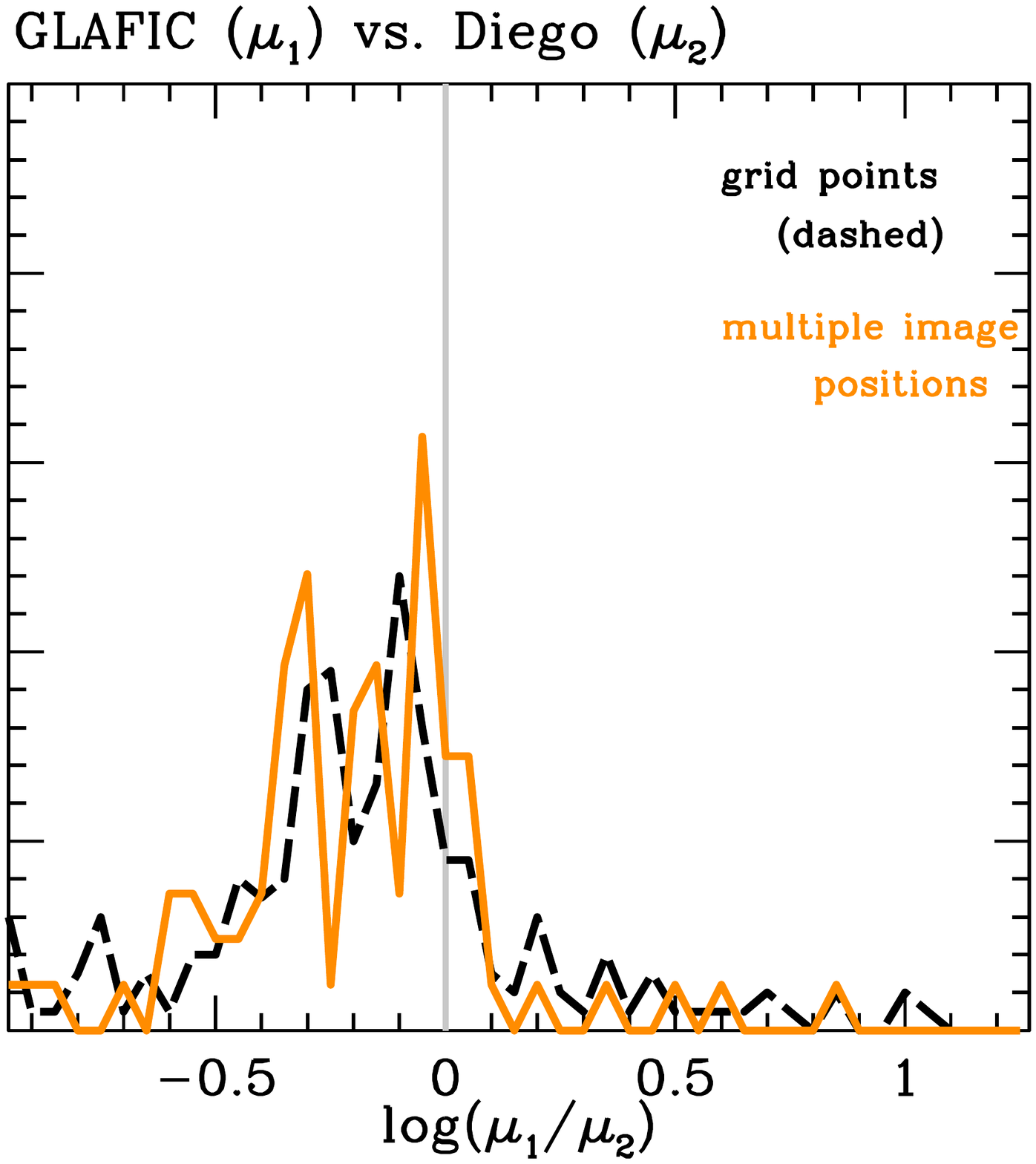}
\vspace{+7pt}
\caption[]{Same as Fig.~\ref{fig:magcorr2744} but for eight pairs of models of MACS J0416.}
\label{fig:magcorr0416}
\end{figure*}

If both models give exactly the same magnifications at all locations, then the histogram of $\log(\mu_1/\mu_2)$ values will be a $\delta$-function at zero. If one model can be obtained from the other by an application of the classic MSD, then the histogram will still be a $\delta$-function, but now displaced from zero. In the language of the MSD, $\log(\mu_1/\mu_2)$ will be $-\log(s^2)$. Note that we use logs, instead of just the ratio $\mu_1/\mu_2$, because interchanging models 1 and 2 simply shifts the $\delta$-function from $-\log(s^2)$ to $\log(s^2)$. Thus, the classic MSD is easy to identify using such histograms. A generalized MSD, such as described in \cite{lie08,lie12}, will, in effect, produce a range of $s$ values across the face of the cluster, and the $\delta$-function histogram will get smeared into a broader peak. This is illustrated in Fig.~\ref{smjori}, where model 2 is the mass distributions shown in fig.~1 of \cite{lie08}, while model 1 was obtained from model 2 by an application of the MSD, and is shown in figs~6 and 7 of that paper. Sources at two different redshifts are present, so the classic MSD is broken. The value of $s$ used in these simulated lenses is 0.75, giving $-\log(s^2)=0.25$ (shown as the red vertical line in our Fig.~\ref{smjori}), which corresponds well to the spike in the black dashed histogram.

Figs~\ref{fig:magcorr2744} and \ref{fig:magcorr0416} present histograms of $\log(\mu_1/\mu_2)$ for the Abell 2744 and MACS J0416 clusters respectively, and for pairs of models, as black dashed lines. In each pair, model 1 is always GLAFIC v3, while models 2 are all the other models, plotted in turn. GLAFIC was chosen as the reference model because it performed best in the \cite{men16} synthetic cluster comparison project.

As in all the previous analyses in this paper we use the same grid of sky locations. It is apparent that MACS J0416 models show very little evidence for the MSD as all black dashed histograms peak at $\log(\mu_1/\mu_2)\approx 0$. The largest deviation from 0 occurs probably in GLAFIC v3 vs. CATS v3, where the typical $\log(\mu_1/\mu_2)\sim -0.15$, which corresponds to $\mu_1/\mu_2\sim 0.7$, or $s\sim 1.19$. The situation is very different for Abell 2744, where some histograms are displaced from zero. The locations of the peaks imply $s$ values between 1.33 and 0.78, so the MSD is not broken in this cluster. Because GLAFIC v3 vs. CATS v3 histogram peaks furthest to the left, while GLAFIC v3 vs. Zitrin-LTM-Gauss v3 peaks furthest to the right, the largest difference in magnifications is between CATS v3 and Zitrin-LTM-Gauss v3, and corresponds to $\log(\mu_1/\mu_2)\sim -0.47$, $\mu_1/\mu_2\sim 0.34$, or $s\sim 1.72$. The reason why the MSD is mostly broken in MACS J0416, but not in Abell 2744 is not clear. It could be because Abell 2744 contains a smaller number of images than MACS J0416 (by about a factor of 2), but other reasons are also possible.

The MSD is most clearly seen in the upper left and upper middle panels of Fig.~\ref{fig:magcorr2744}, in the sense that the black dashed histograms are the most peaked compared to all other similar histograms, and displaced from zero. We speculate that this is because the MSD can operate best between models with similar mass parametrizations. Both GLAFIC and CATS use a superposition of cluster-wide component(s), represented by NFW or PIEMD, and individual galaxies, whose mass properties are tied to their observed magnitude in the visible. In such models, changing the amplitude and/or the scale radius of the cluster-wide potential can mimic the effect of the MSD. One could have expected GLAFIC, Sharon/Johnson and Zitrin-NFW also to show the MSD, but evidently, they do not. The MSD is least pronounced in the bottom right panel of Fig.~\ref{fig:magcorr2744}, which compares a parametric GLAFIC v3 and free-form Williams/\grale~v3, suggesting that the very different parametrizations of the two techniques reduce the effect of global degeneracies. 
However, even here CATS v3 magnification predictions are systematically higher. That CATS v3 and v3.1 predict higher magnifications is already seen in Fig.~\ref{fig:magnifsA2744}. This means that in the source plane, CATS v3 and CATS v3.1 sources will look smaller than those from the other models, but by amounts that will depend on their sky location. 

While MSD and SPT are transformations that apply everywhere within the cluster, including the locations of lensed images, some other degeneracies are local. The most notable among local degeneracies is the monopole degeneracy described in \cite{lie12}, which redistributes mass between images. It is not surprising that there are degeneracies that work {\em around} the images, so to speak, and not {\em at} image locations, because lens modeling uses observed lensed image positions as constraints. 

Therefore it is interesting to ask if the magnifications are better constrained at the locations of the images, where fewer degeneracies are at work, as opposed to other locations. To that end we redid the analysis using the sky locations of the GOLD+SILVER images, shown as red squares in Fig.~\ref{ds9}. These histograms are shown as solid orange lines. Since the number of these images is different for Abell 2744 and MACS J0416, 53 and 88 respectively, and both are different from the number of grid sky locations used for the black dashed histograms, we renormalized all histograms in this section to have the same area under the curve.

In Figs~\ref{fig:magcorr2744} and \ref{fig:magcorr0416}, but especially in the former, the histograms of $\log(\mu_1/\mu_2)$ that use image locations (orange solid lines) show more prominent spikes compared to the histograms that use grid sky locations (black dashed). This is likely because more degeneracies, including monopole degeneracy can act between the locations of the images, while image locations suffer primarily from the MSD. The difference in the presence of the MSD in the two clusters is likely to be the reason why the Normalized Median Absolute Deviation values $\hat\sigma_i$ are larger for Abell 2744 than for MACS J0416. (We note that the displacements between the observed and model predicted image locations, which are typically $\simlt 1^{\prime\prime}$, are too small to account for the magnification differences seen here.)

This section has explored how degeneracies in mass reconstructions impact magnification predictions. For most locations in the multiple image regions of the two clusters we studied, the ratios of high-to-low magnifications of any two models are $\simlt\!2$, as can be seen in Figs~\ref{fig:magcorr2744} and \ref{fig:magcorr0416}, but can rise to significantly higher values over a small fraction of the cluster area. Uncertainties in the magnification translate into those in absolute fluxes of high redshift sources being studied, their volume number density, and hence the luminosity function, i.e. the primary science goal of the HST Frontier Fields project. For example, higher magnification translates into smaller volume at high $z$, higher source intrinsic luminosity, and hence larger luminosity function. (See \cite{bou16} for a detailed discussion.) Proper accounting of the systematic uncertainties in magnifications is critical for the realization of HFF's primary goal of using clusters as accurate cosmic telescopes.

\begin{table*}
\centering
\begin{tabular}{cccccccccc}
\hline Models           & CATS & CATS    & Sharon/  & Zitrin  & Zitrin    & GLAFIC  & Williams&Williams& Diego  \\
	                &      & v3.1    & Johnson  & NFW     & LTM-Gauss &         &         &  v3.1  &        \\
\hline CATS             & -    &0.92|3.06& 9.83|5.26&5.08|3.29&14.72|15.75&7.66|5.00&2.74|1.51&---|2.00&---|2.47\\
\hline CATS v3.1        &      &  -      &16.08|5.20&6.06|3.41&31.62|20.33&12.02|4.42&3.291|1.68&---|2.15&---|2.76\\ 
\hline Sharon           &      &         &   -      &3.12|2.44& 6.21| 9.87&2.75|2.70&1.46|1.43&---|2.16&---|2.99\\ 
\hline Zitrin NFW       &      &         &          &   -     & 7.12| 5.63&1.64|2.48&1.50|1.19&---|2.01&---|2.27\\ 
\hline Zitrin LTM-Gauss &      &         &          &         &    -      &6.67|6.67&1.45|1.63&---|1.78&---|3.03\\ 
\hline GLAFIC           &      &         &          &         &           & -       &1.51|1.72&---|2.12&---|2.72\\  
\hline Williams         &      &         &          &         &           &         &   -     &---|0.88&---|1.11\\  
\hline Williams v3.1    &      &         &          &         &           &         &         &   -    &---|1.2 \\  
\hline Diego            &      &         &          &         &           &         &         &        & -      \\  
\hline 
\end{tabular} 
\caption{The median difference between two models 1 and 2, normalized by the average of the two models' quoted uncertainties: ${\rm median}\Bigl[(\mu_1-\mu_2)/[0.5(\delta_1+\delta_2)]\Bigr]$, where the median is taken over $N=214$ sky locations, and the ordering of 1 and 2, and the use of upper vs. lower error-bars depends on which of the two magnifications is larger at a given sky location. The $\delta_1$ and $\delta_2$ error-bars are the 68\% confidence limits. All models are v3 unless stated otherwise. The two values on either side of the vertical bar (in each cell in the table) are for Abell 2744 and MACS J0416, respectively.}
\label{table3}
\end{table*}

\section{Conclusions}
\label{sec:con}

The comparison of models presented here provides a birds-eye view of the state of the art of cluster lensing reconstruction. We show that despite the overall agreement among models, the differences are not insignificant, and often exceed the statistical errors of individual reconstructions. We have argued that the most important source for the differences in predicted magnifications are lensing degeneracies. While the classic global mass sheet degeneracy is broken by sources at multiple redshifts, the generalized version of that degeneracy (where the scaling of the mass and the added mass sheet vary somewhat across the face of the cluster), as well as other, local degeneracies are alive and well. A corollary is that a small lens plane rms is not a sufficient condition for the model to be a good representation of the true magnification distribution of the cluster. Despite degeneracies, the total cluster mass is likely to be well constrained by all models. Degeneracies redistribute mass within the lens, but have little affect on the total mass, which is an integral quantity, largely determined by the radial extent of the lensed images from the cluster centre.

Model comparison of the type presented here can be used to determine how well the data constrains the mass and magnification map of a cluster: if there is a significant dispersion between models' predictions, then a cluster is not well constrained, despite what the statistical uncertainties from individual models may imply. In the present case, MACS J0416 is constrained considerably better than Abell 2744, whose models suffer from the generalized mass sheet degeneracy, and possibly from the source-plane transformation (Section~\ref{degen}). The reason why the MSD is not an issue in MACS J0416, but is present among models of Abell 2744 is not clear. One possibility is that MACS J0416 has more multiple images, thereby restricting the MSD. Quality of lensing data, like accurate redshifts, may play a role,  however, MSD-degenerate models of clusters with perfectly known data are easy to construct \citep{lie08}. 

We show that the same set of models (for example, the 7 models that are in common between Abell 2744 and MACS J0416) can perform differently in two clusters. For example, in one cluster, a given model may consistently over-predict the magnifications, while in another, its predictions could track the median very well. This conclusion has implications for model testing on simulated data. Even if a given model performs well in tests where a synthetic cluster is being reconstructed, one must be careful in claiming that method to be superior to others: the method may work well on clusters generated using certain prescriptions, but not necessarily on real ones, whose physics may differ from that assumed in simulations. This makes the analysis presented in this paper, which does not rely on synthetic clusters, an essential part of model comparison.

For the v3 and v3.1 round of HFF reconstructions, the number of submitted parametric models far outnumbered free-form ones: Williams/\grale~ is fully free-form, and Diego is hybrid. The rest are parametric, and 3 (CATS v3, v3.1, and Sharon/Johnson) use the same modeling software, \lenstool. Because of that, we cannot currently assess the differences in the performance of the two types of reconstructions: parametric vs. free-form, but hope to do so in the future, when more models become available. Another interesting avenue is to explore the differences between v1 and v3 models. Preliminary examination suggests that v1 models of Abell 2744 were in better agreement with each other than is the case with v3 models.

In this study we represented the actual magnifications---which are unknown---with the median values from all models. While there is no guarantee that the median is the true magnification, it is the best guess. In the future it will be interesting to repeat this analysis with synthetic clusters instead of observed ones. This will provide information both about how well a single model's magnifications match the true ones, as well as how well the median of all models, $\tilde{\mu}$, tracks this true magnification. The remark made earlier stays relevant of course: while it is certainly a good idea to benchmark different models against simulated clusters, one must avoid the pitfall of declaring one model to be superior simply because it performs the best on simulated data.

\section*{Acknowledgments}

We are grateful to all those who contributed to the data used here: 
Anahita Alavi, 
Marusa Bradac, 
Gabe Brammer, 
Benjamin Clement, 
Jose Diego, 
Claudio Grillo, 
Austin Hoag, 
Jens Hjorth,
Mathilde Jauzac, 
Traci Johnson, 
Ryota Kawamata, 
Jean-Paul Kneib, 
Daniel Lam, 
Priya Natarajan, 
Masamune Oguri
Johan Richard, 
Kevin Sebesta,
Jonatan Selsing,
Irene Sendra, 
Keren Sharon, 
Brian Siana, 
Tomasso Treu, 
Xin Wang, and
Adi Zitrin.
We would like to thank Minnesota Supercomputing Institute whose technical support and computer resources were invaluable in producing the \grale~ set of models. Financial support for this work was provided to S.A.R. by NASA through grant HST-GO-13386 from the Space Telescope Science Institute (STScI), which is operated by Associated Universities for Research in Astronomy, Inc. (AURA), under NASA contract NAS 5-26555. The background HFF images of Abell 2744 and MACS J0416 in Fig.~\ref{ds9} were taken from {\tt https://www.spacetelescope.org/images/heic1111c} and {\tt https://frontierfields.org/meet-the-frontier-fields/}, respectively. We thank the referee for the careful reading of the manuscript and suggestions that helped improve the paper.

% Don't change these lines
\bsp	% typesetting comment
\label{lastpage}

\begin{thebibliography}{99}

\bibitem[Amanullah et al.(2011)]{ama11}
Amanullah, R. et al. 2011, ApJ, 742, L7
%A Highly Magnified Supernova at z = 1.703 behind the Massive Galaxy Cluster A1689

\bibitem[Atek et al.(2015)]{ate15}
Atek, H. et al. 2015, ApJ, 814, 69
%Are Ultra-faint Galaxies at z = 6-8 Responsible for Cosmic Reionization? Combined Constraints from the Hubble Frontier Fields Clusters and Parallels
	
\bibitem[Blandford(1990)]{bla90}
Blandford, R.D. 1990, QJRAS, 31, 305
%GL

\bibitem[Bouwens et al.(2016)]{bou16}
Bouwens, R.J., Oesch, P.A., Illingworth, G.D., Ellis, R.S. \& Stefanon, M. Preprint, arXiv:1610.00283
%THE Z~6 LUMINOSITY FUNCTION FAINTER THAN −15 MAG FROM THE HUBBLE FRONTIER FIELDS: THE IMPACT OF MAGNIFICATION UNCERTAINTIES

\bibitem[Broadhurst et al.(2005)]{br05}
Broadhurst et al. 2005, ApJ, 621, 53.

\bibitem[Coe et al.(2015)]{coe15}
Coe, D., Bradley, L. \& Zitrin, A. 2015, ApJ, 800, 84
Frontier Fields: High-redshift Predictions and Early Results

\bibitem[Diego et al.(2015)]{die15}
Diego, J., Broadhurst, T., Molnar, S.M., Lam, D. \& Lim, J. 2015, MNRAS, 447, 3130
%Free-form lensing implications for the collision of dark matter and gas in the frontier fields cluster MACS J0416.1-2403

\bibitem[Finkelstein et al.(2015)]{fin15}
Finkelstein, S.L. et al. 2015, ApJ, 810, 71
%The Evolution of the Galaxy Rest-frame Ultraviolet Luminosity Function over the First Two Billion Years

\bibitem[Gorenstein et al.(1988)]{gsf88}
Gorenstein, M. V., Shapiro, I. I. \& Falco, E. E. 1988, ApJ, 327, 693
%Degeneracies in parameter estimates for models of gravitational lens systems

\bibitem[Hainline et al.(2009)]{hai09}
Hainline, K. N., Shapley, A. E., Kornei, K. A., Pettini, M., Buckley-Geer, E., Allam, S. S. \& Tucker, D. L. 2009, ApJ, 701, 52
%Rest-Frame Optical Spectra of Three Strongly Lensed Galaxies at z ~ 2

\bibitem[Hammer(1990)]{ham90}
Hammer, F. 1990, Ap\&SS, 170, 389 
%New views through the gravitational telescope

\bibitem[Jauzac et al.(2014)]{jau14}
Jauzac, M. et al. 2014, MNRAS, 443, 1549
%Hubble Frontier Fields: a high-precision strong-lensing analysis of galaxy cluster MACSJ0416.1-2403 using ˜200 multiple images

\bibitem[Jauzac et al.(2015)]{jau15}
Jauzac, M. et al. 2015, MNRAS, 446, 4132
%Hubble Frontier Fields: the geometry and dynamics of the massive galaxy cluster merger MACSJ0416.1-2403

\bibitem[Johnson et al.(2014)]{john14}
Johnson, T., Sharon, K., Bayliss, M., Gladders, M., Coe, D. \& Eeling, H. 2014, ApJ, 797, 48
%Lens Models and Magnification Maps of the Six Hubble Frontier Fields Clusters

\bibitem[Jullo \& Kneib(2009)]{jk09}
Jullo, E. \& Kneib, J.-P., 2009. MNRAS, 395, 1319
%Multiscale cluster lens mass mapping - I. Strong lensing modelling

\bibitem[Jullo et al.(2007)]{j07}
Jullo, E., Kneib, J.-P., Limousin, M., Eliasdottir, A., Marshall, P.J. \& Verdugo, T. 2007, New Journal of Physics, 9, 447
%A Bayesian approach to strong lensing modelling of galaxy clusters

\bibitem[Kassiola \& Kovner(1993)]{kk93}
Kassiola, A. \& Kovner, I. 1993, ApJ, 417, 450
%Elliptic Mass Distributions versus Elliptic Potentials in Gravitational Lenses

\bibitem[Kawamata et al.(2016)]{kaw16}
Kawamata, R., Oguri, M., Ishigaki, M., Shimasaku, K. \& Ouchi, M. 2016, ApJ, 819, 114    %% arXiv:1510.06400
%Precise Strong Lensing Mass Modeling of Four Hubble Frontier Field Clusters and a Sample of Magnified High-redshift Galaxies

\bibitem[Kawamata et al.(2015)]{kaw15}
Kawamata, R., Ishigaki, M., Shimasaku, K., Oguri, M. \& Ouchi, M. 2015, ApJ, 804, 103
%The Sizes of z ˜ 6-8 Lensed Galaxies from the Hubble Frontier Fields Abell 2744 Data

\bibitem[Keeton(2001)]{kee01}
Keeton, C.  2001, arXiv:astro-ph/0102341
%A Catalog of Mass Models for Gravitational Lensing

\bibitem[Kelly et al.(2016)]{kel16}
Kelly, P.L. et al. 2016, ApJ, 819, L8
%Deja Vu All Over Again: The Reappearance of Supernova Refsdal

\bibitem[Kelly et al.(2015)]{kel15}
Kelly, P.L. et al. 2015, Science, 347, 1123
%Multiple images of a highly magnified supernova formed by an early-type cluster galaxy lens

\bibitem[Lam et al.(2014)]{lam14}
Lam, D., Broadhurst, T., Diego, J.M., Lim, J., Coe, D., Ford, H.C. \& Zheng, W. 2014, ApJ, 797, 98
%A Rigorous Free-form Lens Model of A2744 to Meet the Hubble Frontier Fields Challenge

\bibitem[Laporte et al.(2016)]{lap16}
Laporte, N. et al. 2016, ApJ, 820, 98 %% arXiv:1602.02775
%Young Galaxy Candidates in the Hubble Frontier Fields - III. MACSJ0717.5+3745

\bibitem[Liesenborgs et al.(2006)]{lie06}
Liesenborgs, J., De Rijcke, S. \& Dejonghe, H. 2006, MNRAS, 367, 1209
%A genetic algorithm for the non-parametric inversion of strong lensing systems

\bibitem[Liesenborgs et al.(2007)]{lie07}
Liesenborgs, J., De Rijcke, S., Dejonghe, H. \& Bekaert, P. 2007, MNRAS, 380, 1729
%Non-parametric inversion of gravitational lensing systems with few images using a multi-objective genetic algorithm

\bibitem[Liesenborgs et al.(2008)]{lie08}
Liesenborgs, J., De Rijcke, S., Dejonghe, H., Bekaert, P. 2008, MNRAS, 386, 307
%A generalization of the mass-sheet degeneracy producing ring-like artefacts in the lens mass distribution

%\bibitem[Liesenborgs et al.(2009)]{lie09}
%Liesenborgs, J., de Rijcke, S., Dejonghe, H., Bekaert, P. 2009, MNRAS, 397, 341
%Non-parametric strong lens inversion of SDSS J1004+4112

\bibitem[Liesenborgs et al.(2012)]{lie12}
Liesenborgs, J. \& De Rijcke, S. 2012, MNRAS, 425, 1772
%Lensing degeneracies and mass substructure

\bibitem[Limousin et al.(2016)]{lim16}
Limousin, M. et al. 2016, A\&A, 588, 99
%Strong-lensing analysis of MACS J0717.5+3745 from Hubble Frontier Fields observations: How well can the mass distribution be constrained?

\bibitem[Masafumi et al.(2015)]{mas14}
Masafumi, I., Ryota, K., Masami, O., Masamune, O., Kazuhir, S. \& Yoshiaki, O. 2015, ApJ, 799, 12   %% arXiv:1408.6903
%Hubble Frontier Fields First Complete Cluster Data: Faint Galaxies at z∼5−10 for UV Luminosity Functions and Cosmic Reionization

\bibitem[McLeod et al.(2015)]{mcl15}
McLeod, D.J., McLure, R.J., Dunlop, J.S., Robertson, B.E., Ellis, R.S. \& Targett, T.A. 2015, MNRAS, 450, 3032
%New redshift z ≃ 9 galaxies in the Hubble Frontier Fields: implications for early evolution of the UV luminosity density

\bibitem[Meneghetti et al.(2016)]{men16}
Meneghetti, M. et al., 2016, MNRAS, Preprint, arXiv:1606.04548
%The Frontier Fields Lens Modeling Comparison Project

%\bibitem[Menghetti et al.(2014)]{men14}
%Menghetti, M. et al., 2014, ApJ, 797, 34

\bibitem[Navarro, Frenk \& White(1997)]{nfw97}
Navarro, J. F., Frenk, C. S. \& White, S. D. M. 1997, ApJ, 490, 493

\bibitem[Nordin et al.(2014)]{nor14}
Nordin, J. et al. 2014, MNRAS, 440, 2742
%Lensed Type Ia supernovae as probes of cluster mass models

\bibitem[Ogrean et al.(2015)]{ogr15}
Ogrean, G.A. et al. 2015, ApJ, 812, 153
%Frontier Fields Clusters: Chandra and JVLA View of the Pre-merging Cluster MACS J0416.1-2403

\bibitem[Oguri(2010)]{ogu10}
Oguri, M. 2010, PASJ, 62, 1017

\bibitem[Patel et al.(2014)]{pat14}
Patel, B. et al. 2014, ApJ, 786, 9
%Three Gravitationally Lensed Supernovae behind CLASH Galaxy Clusters

\bibitem[Pettini et al.(2000)]{pet00}
Pettini, M., Steidel, C. C., Adelberger, K. L., Dickinson, M. \& Giavalisco, M. 2000, ApJ, 528, 96
%The Ultraviolet Spectrum of MS 1512-CB58: An Insight into Lyman-Break Galaxies

\bibitem[Richard et al.(2014)]{ric14}  %arXiv:1405.3303
Richard, J. et al. 2014, MNRAS, 444, 268
%Mass and magnification maps for the Hubble Space Telescope Frontier Fields clusters: implications for high-redshift studies

\bibitem[Riehm et al.(2011)]{rie11}
Riehm, T. et al. 2011, A\&A 536, A94
%Near-IR search for lensed supernovae behind galaxy clusters III. Implications for cluster modeling and cosmology

\bibitem[Rodney et al.(2016)]{rod16} 
Rodney, S. A. et al. 2016, ApJ, 820, 50
%SN Refsdal: Photometry and Time Delay Measurements of the First Einstein Cross Supernova

\bibitem[Rodney et al.(2015)]{Rodney}  %arXiv:1505.06211
Rodney, S. A. et al. 2015, ApJ, 811, 70
%Illuminating a Dark Lens : A Type Ia Supernova Magnified by the Frontier Fields Galaxy Cluster Abell 2744. 

\bibitem[Saha(2000)]{saha00}
Saha, P., 2000, AJ, 120, 1654
%Lensing Degeneracies Revisited

\bibitem[Schechter(1976)]{sch76}
Schechter, P. 1976, ApJ, 203, 297
%An analytic expression for the luminosity function for galaxies.

\bibitem[Schneider \& Sluse(2014)]{ss14}
Schneider, P. \& Sluse, D. 2014, A\&A, 564, 103
%Source-position transformation: an approximate invariance in strong gravitational lensing

\bibitem[Schneider \& Sluse(2013)]{ss13}
Schneider, P. \& Sluse, D. 2013, A\&A, 559, 37
%Mass-sheet degeneracy, power-law models and external convergence: Impact on the determination of the Hubble constant from gravitational lensing

\bibitem[Sebesta et al.(2016)]{seb16}
Sebesta, K., Williams, L.L.R., Mohammed, I., Saha, P. \& Liesenborgs, J. 2016, MNRAS, 461, 2126 
%Testing light-traces-mass in Hubble Frontier Fields Cluster MACS-J0416.1-2403

\bibitem[Swinbank et al.(2007)]{swi07}
Swinbank, A. M., Bower, R. G., Smith, G. P., Wilman, R. J., Smail, I., Ellis, R. S., Morris, S. L. \& Kneib, J.-P. 2007, MNRAS, 376, 479
%Resolved spectroscopy of a gravitationally lensed L* Lyman-break galaxy at z ~ 5

\bibitem[Treu et al.(2016)]{tre16}
Treu, T. et al. 2016, ApJ, 817, 60
%"Refsdal" Meets Popper: Comparing Predictions of the Re-appearance of the Multiply Imaged Supernova Behind MACSJ1149.5+2223

\bibitem[Unruh et al.(2016)]{unr16}
Unruh, S., Schneider, P. \& Sluse, D. 2016, Preprint, arXiv:1606.04321
%Ambiguities in gravitational lens models: the density field from the source position transformation

\bibitem[Wang et a.(2015)]{wan15}
Wang, X. et al. 2015, ApJ, 811, 29
%The Grism Lens-amplified Survey from Space (GLASS). IV. Mass Reconstruction of the Lensing Cluster Abell 2744 from Frontier Field Imaging and GLASS Spectroscopy

\bibitem[Zitrin et al.(2013)]{zit13}
Zitrin et al. 2013, ApJ, 762, L30

\bibitem[Zitrin et al.(2009)]{zit09}
Zitrin et al. 2009, MNRAS, 396, 1985

\end{thebibliography}
\end{document}